\documentclass[final,5p,times,twocolumn]{elsarticle} 

\DeclareMathAlphabet{\mathpzc}{OT1}{pzc}{m}{it}
\usepackage{lineno}
\usepackage{setspace}
\usepackage{epstopdf}
\usepackage{graphicx}
\usepackage{amssymb}
\usepackage{amsmath}
\usepackage{amsthm}
\usepackage[colorlinks=true]{hyperref}
\usepackage{breakurl}
\usepackage[all]{hypcap}
\usepackage{mathrsfs}

\usepackage[percent]{overpic}
\usepackage{epsfig}
\usepackage{verbatim}
\usepackage{multirow}
\usepackage{float}
\usepackage{array}
\usepackage[T1]{fontenc}
\usepackage{url}
\usepackage{hyperref}
\usepackage{natbib}
\usepackage{wrapfig}
\usepackage{amsmath}
\usepackage{amsmath}

\journal{Elsevier}
\begin{document}

\begin{frontmatter}
\title{Design, simulation and performance of the resistive-anode PICOSEC Micromegas detector}

\author[1]{D. Janssens}
\author[2]{A. Utrobicic\corref{cor1}}\ead{antonija.utrobicic@irb.hr}
\author[3]{M. Kovacic}
\author[1]{M. Lisowska}

\author[4]{R. Aleksan}
\author[5]{Y. Angelis}
\author[6]{J. Bortfeldt}
\author[1]{F. Brunbauer}
\author[7,8]{M. Brunoldi}
\author[2]{M. Buljan}
\author[5]{E. Chatzianagnostou}
\author[9]{J. Datta}
\author[1]{R. De Oliveira}
\author[10]{G. Fanourakis}
\author[7,8]{D. Fiorina\fnref{fn1}}
\author[1]{K. J. Floethner}
\author[11]{M. Gallinaro}
\author[12]{F. Garcia}
\author[4]{I. Giomataris}
\author[13]{K. Gnanvo}
\author[1]{A. Gurpinar}
\author[4]{F.J. Iguaz\fnref{fn2}}
\author[4]{A. Kallitsopoulou}
\author[5]{I. Karakoulias }
\author[13]{B. Kross}
\author[4]{P. Legou}
\author[14]{J. Liu}
\author[15,16]{M. Lupberger}
\author[1,5]{I. Maniatis\fnref{fn3}}
\author[13]{J. McKisson}
\author[14]{Y. Meng}
\author[1,16]{H. Muller}
\author[1]{E. Oliveri}
\author[17]{G. Orlandini}
\author[13]{A. Pandey}
\author[4]{T. Papaevangelou}
\author[18]{M. Pomorski}
\author[1]{L. Ropelewski}
\author[2]{K. Salamon}
\author[5,19]{D. Sampsonidis}
\author[1]{L. Scharenberg}
\author[1]{T. Schneider}
\author[4]{E. Scorsone}
\author[4]{L. Sohl \fnref{fn4}}
\author[20]{Y. Tsipolitis}
\author[5,19]{S.E. Tzamarias}
\author[7,8]{I. Vai}
\author[1]{M. van Stenis}
\author[1]{R. Veenhof}
\author[7,8]{P. Vitulo}
\author[14]{X. Wang}
\author[11]{S. White}
\author[13]{W. Xi}
\author[14]{Z. Zhang}
\author[14]{and Y. Zhou}

\address[1]{European Organization for Nuclear Research (CERN), CH-1211, Geneve 23, Switzerland}
\address[2]{Ruđer Bošković Institute, Bijenička cesta 54, 10000 Zagreb, Croatia}
\address[3]{University of Zagreb, Faculty of Electrical Engineering and Computing, 10000 Zagreb, Croatia}
\address[4]{IRFU, CEA, Université Paris-Saclay, F-91191 Gif-sur-Yvette, France}
\address[5]{Department of Physics, Aristotle University of Thessaloniki, University Campus, GR-54124, Thessaloniki, Greece}
\address[6]{Department for Medical Physics, Ludwig Maximilian University of Munich, Am Coulombwall 1, 85748 Garching, Germany}
\address[7]{Dipartimento di Fisica, Università di Pavia, Via Bassi 6, 27100 Pavia (IT)}
\address[8]{INFN Sezione di Pavia, Via Bassi 6, 27100 Pavia (IT)}
\address[9]{Stony Brook University, Dept. of Physics and Astronomy, Stony Brook, NY 11794-3800, USA}
\address[10]{Institute of Nuclear and Particle Physics, NCSR Demokritos, GR-15341 Agia Paraskevi, Attiki, Greece}
\address[11]{Laboratório de Instrumentação e Física Experimental de Partículas, Lisbon, Portugal}
\address[12]{Helsinki Institute of Physics, University of Helsinki, FI-00014 Helsinki, Finland}
\address[13]{Jefferson Lab, 12000 Jefferson Avenue, Newport News, VA 23606, USA}
\address[14]{State Key Laboratory of Particle Detection and Electronics, University of Science and Technology of China, Hefei 230026, China}
\address[15]{Helmholtz-Institut für Strahlen- und Kernphysik, University of Bonn, Nußallee 14–16, 53115 Bonn, Germany}
\address[16]{Physikalisches Institut, University of Bonn, Nußallee 12, 53115 Bonn, Germany}
\address[17]{Friedrich-Alexander-Universität Erlangen-Nürnberg, Schloßplatz 4, 91054 Erlangen, Germany}
\address[18]{CEA-LIST, Diamond Sensors Laboratory, CEA Saclay, F-91191 Gif-sur-Yvette, France}
\address[19]{Center for Interdisciplinary Research and Innovation (CIRI-AUTH), Thessaloniki 57001, Greece}
\address[20]{National Technical University of Athens, Athens, Greece}

\cortext[cor1]{Correspondence to: Ru\dj er Bo\v{s}kovi\'{c} Institute,  Bijeni\v{c}ka cesta 54, 10000 Zagreb, Croatia}

\fntext[fn1]{Now at Gran Sasso Science Institute, Viale F. Crispi, 7 67100 L'Aquila, Italy}
\fntext[fn2]{Now at SOLEIL Synchrotron, L’Orme des Merisiers, Départementale 128, 91190 Saint-Aubin, France.}
\fntext[fn3]{Now at Department of Particle Physics and Astronomy, Weizmann Institute of Science, Rehovot, 7610001, Israel.}
\fntext[fn4]{Now at TÜV NORD EnSys GmbH Co. KG.}

\begin{abstract}
The PICOSEC Micromegas detector is a Micro-Pattern Gaseous Detector (MPGD) concept developed to achieve tens of picosecond-level timing resolution for charged particle detection by combining a Cherenkov radiator with a two-stage Micromegas amplification structure. To improve operational robustness at high gain and under intense radiation backgrounds, a resistive anode has been implemented using a diamond-like carbon (DLC) layer deposited on a Kapton substrate. While this design enhances detector stability, the resistive layer may influence rate capability, signal formation, and detector capacitance, altering its timing performance.\\

In this work, a comprehensive study of a resistive PICOSEC design is presented, including an analytical model and finite-element simulation to quantify rate-dependent gain reduction due to ohmic voltage drop in the resistive layer. An analytical solution for the voltage distribution across a finite-size resistive layer is derived, and a numerical model is developed to evaluate gain suppression under intense particle fluxes. For the single-channel prototype geometry and expected beam conditions at the CERN SPS H4 beam line, surface resistivities around 20 M$\Omega/\Box$ are found to ensure discharge protection and acceptable gain stability. The impact of the resistive layer on signal integrity is investigated using an extended Ramo–Shockley formalism with time-dependent weighting fields and Garfield++ simulations. The contribution of delayed signal components induced by the resistive layer is quantified, and a preservation of the leading-edge of the signal was found for surface resistivities exceeding 100 k$\Omega/\Box$. \\

Single-channel resistive-anode prototypes were designed ($\varnothing$10 and $\varnothing$15 mm), constructed, and experimentally characterized. Laboratory measurements using single photoelectrons and a power spectral density analysis show the predicted reduction in signal amplitude due to the insulating layer, while preserving the leading edge of the electron peak. Muon beam tests with both CsI and DLC photocathodes were performed, demonstrate a time resolution of $11.5\pm 0.4$ ps using CsI, comparable to the $11.9\pm0.4$ ps of the metallic-anode device, showing the suitability of the resistive design for precision timing applications in challenging operational conditions.
\end{abstract}
\end{frontmatter}

\newpage
\section{Introduction}
\label{sec:chap1}
\noindent The increasing collision densities foreseen in future high-luminosity colliders pose unprecedented challenges for track and vertex reconstruction, with up to about 200 simultaneous proton–proton interactions expected per bunch crossing at the High-Luminosity Large Hadron Collider. Preserving efficient event separation under these pile-up conditions requires detector systems with $\mathcal{O}(10)$ ps timing resolution for minimum ionizing particles, while also tolerating intense particle fluxes and high radiation levels \cite{Detector:2784893}. By incorporating precision timing for track reconstruction, a time-layered approach can be used, which dramatically mitigates pile-up, allowing collisions that overlap in space to be distinguished by their differing time of occurrence. Micro-Pattern Gaseous Detectors (MPGDs) have proven to be robust and scalable \cite{6551204}, offering notable rate capability \cite{poli_lener_2005_k16h7-bjh83}, exhibit good spatial resolution \cite{KETZER2004314}, and provide a cost-effective solution for instrumenting large detector areas \cite{Ziegler:2002yc, ABBANEO2013383, Kawamoto:1552862, CERN-LHCC-2013-020}. However, reaching sub-nanosecond timing precision with gas-based detectors is inherently challenging. The intrinsic timing spread of MPGD signals is limited by the Poissonian fluctuations of the primary cluster position in the gas and the diffusion effects during electron drift. To overcome these limitations, different designs have been proposed to minimize the time jitter \cite{CHARPAK199163,DERRE2000314,BRESKIN2002670}, yet had difficulty reaching timing precisions of less then $\mathcal{O}(100)$ ps. \\

\begin{figure}[!t]
\begin{center}

\includegraphics[width=0.93\columnwidth]{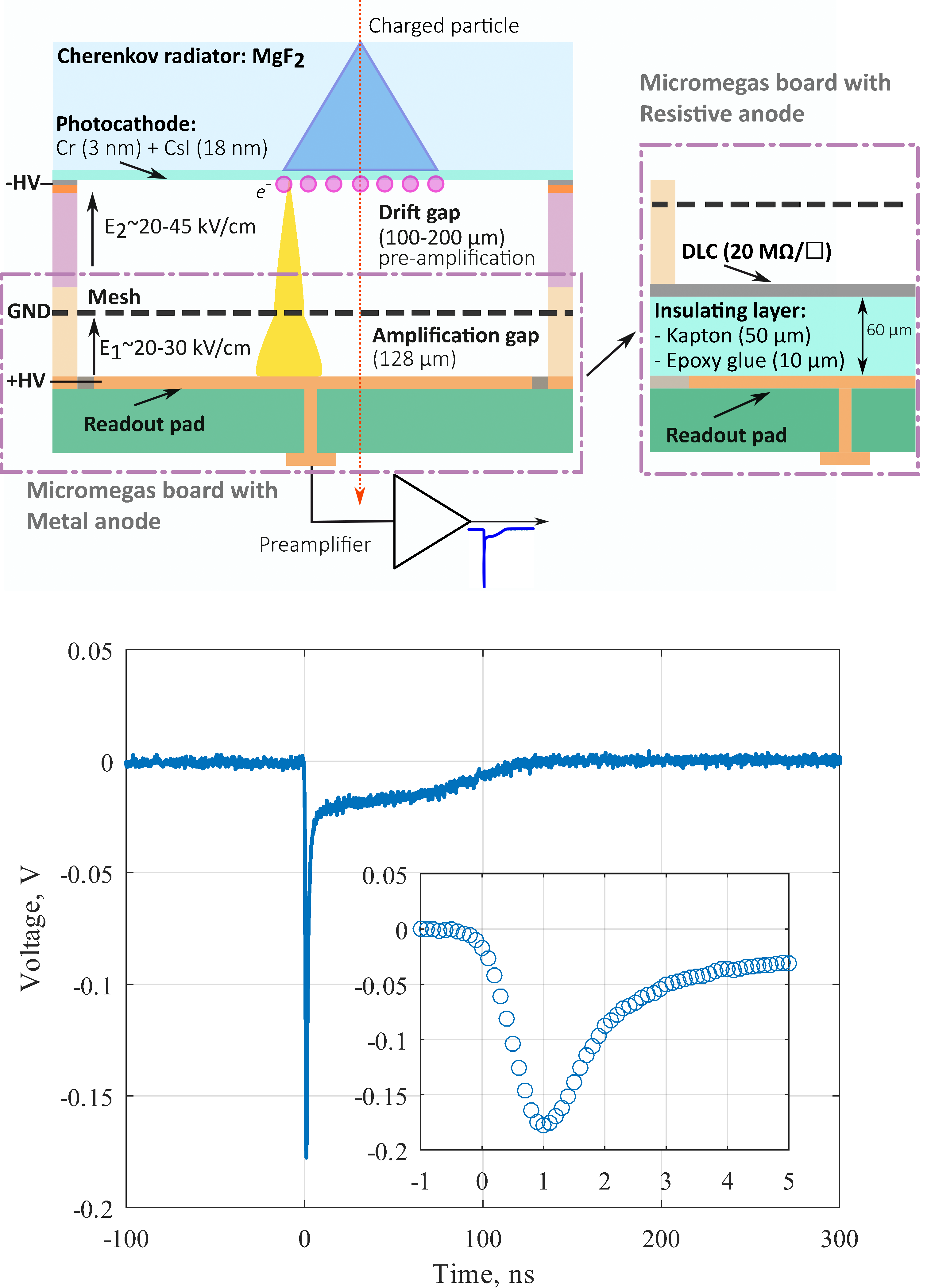}

\caption{Top: Operating principle of the PICOSEC MM detector. Bottom: Typical PICOSEC MM signal at the output of the amplifier.}
\label{fig_concept}
\end{center}
\end{figure}

The PICOSEC Micromegas (hereafter PICOSEC MM) is a precise timing MPGD developed to achieve a $\mathcal{O}(10)$ ps time resolution for relativistic charged particles. This performance is realized through the implementation of a two-stage Micromegas amplification structure \cite{Giomataris1996}, coupled to a Cherenkov radiator that also serves as the detector’s entrance window. The radiator, typically a MgF$_2$ crystal, is coated with a semi-transparent photocathode such as CsI \cite{bortfeldt2018picosec,Sohl2020}. A schematic overview of the detector layout and its operating principle is shown in the top panel of Fig.~\ref{fig_concept}. When a relativistic charged particle traverses the Cherenkov radiator, ultraviolet (UV) photons are emitted and subsequently converted into primary photoelectrons by the photocathode over the area covered by the Cherenkov cone. Under a  $\approx$35 kV/cm electric field, the emitted photoelectrons promptly initiate a Townsend avalanche within the gas volume, reducing the observed time jitter. The amplification process takes place in two distinct regions separated by a woven micro-mesh embedded in the bulk Micromegas \cite{GIOMATARIS2006405}: a pre-amplification region of approximately 200 µm above the mesh, and a nominal 128 µm amplification gap mechanically defined by pillars between the mesh and the anode plane. The detector operates at atmospheric pressure and room temperature using a gas mixture of 80\% Ne, 10\% CF$_4$, and 10\% C$_2$H$_6$, hereafter referred to as the COMPASS gas mixture\footnote{While being the current baseline mixture it has a relatively high global warming potential of 740 (normalized to CO$_2$) compared to mixtures such as Ar/CO$_2$ (93\%/7\%) that has 0.07. There is an ongoing effort to look for an alternative gas mixture that can provide a good timing performance \cite{AIME2024169903}}. Once electrons traverse the mesh, their motion induces a fast current signal on the readout electrodes. A representative waveform is shown in the bottom Fig.~\ref{fig_concept}, where the leading edge of the electron peak constitutes the primary timing observable.\\

However, operating Micromegas detectors at the high gains of $\mathcal{O}\left(10^5-10^6\right)$ required for optimal timing performance, or withing highly ionizing radiation backgrounds, poses the risk of electrical discharges that can damage the device. To mitigate this, a resistive anode has been introduced in the PICOSEC design. As depicted in the inset of the top Fig. \ref{fig_concept}, the resistive anode is implemented as a thin diamond-like carbon (DLC) film coated on an insulating Kapton laminate separating the anode from the readout plane. Such a resistive layer offers increased stability: in the event of a spark or excessive current, its high sheet resistance limits the instantaneous current flow, thereby quenching discharges and preventing damage to the detector components. This approach follows the general principle adopted in other resistive MPGDs, e.g., resistive Micromegas \cite{ALEXOPOULOS2011110} and µ-Resistive Well (µRWELL) \cite{Bencivenni_2015} designs, where a surface resistivity of at least 10 M$\Omega/\Box$ is used to ensure minimal spark protection. However, besides improving the stability against discharges and separating the high voltage from the readout electronics, the resistive anode can adversely impact the detector’s rate capability. At very high event rates, the flow of charge in a resistive electrode can produce localized voltage drops (ohmic polarization) that temporarily reduce the effective amplification field. This rate-dependent gain suppression can limit the detector’s efficiency and timing performance if the resistivity and grounding scheme are not optimized. Moreover, the presence of a resistive layer influences the signal formation on the readout: the induced current is not only from the movement of the election-ions pairs in the amplification region but also the reaction of the resistive material due to its finite conductivity. This can result in an alteration of the pulse shape as seen by the front-end electronics by the addition of a delayed signal that, given sufficiently low surface resistivities, can be shared over neighboring readout channels. In addition, the detector capacitance as seen by the electronics is modified, which alters the shaping of the electron peak \cite{utrobicic2025single}. Furthermore, the resistive layer acts as a source of Johnson–Nyquist thermal noise~\cite{Einstein1906,Johnson1928,Nyquist1928} in the system, which was treated numerically in Ref.~\cite{JANSSENS2024169374}. These signal integrity considerations must be addressed to ensure that the leading edge of the signal pulse is minimally affected, thereby preserving the timing capability of the device.\\

With the aim of developing a robust multi-channel implementation of the PICOSEC MM detector, a new design based on the resistive readout structure shown in the top Fig. \ref{fig_concept} has been developed. In this work, a comprehensive study of the resistive PICOSEC MM design is presented, combining analytical modeling, finite-element simulations, and experimental characterization. The rate capability is estimated through analytical and numerical modeling of the voltage drop in the resistive layer under intense irradiation, allowing the estimation of gain reduction as a function of surface resistivity and particle flux. In parallel, the impact of the resistive layer on signal formation is studied using an extended Ramo-Shockley formalism with time-dependent weighting fields and Garfield++ simulations, quantifying the contribution of delayed signal components and their influence on the leading edge of the electron-peak. Two single-channel resistive-anode prototypes with $\mathcal{O}(10)$ mm active areas were designed, constructed, and characterized. Laboratory measurements using single photoelectrons were performed and compared with their metal-anode counterparts in order to study the predicted signal-amplitude reduction due to the insulating layer and to assess the signal integrity of the electron-peak leading edge. Finally, the timing performance was evaluated in a 150~GeV muon beam at CERN using both CsI and DLC photocathodes, and directly compared with a metallic-anode detector operated under identical conditions.



\section{Simulations}

\subsection{Rate capability}
\noindent Because of its non-zero resistance to ground, the resistive anode causes an ohmic reduction of the electric field within the amplification gap when high gains or high event rates are encountered. The decrease in gain that follows adversely affects the gas gain performance, the degree of which is dependent on the surface resistivity $R$ [$\Omega/\Box$] and precise grounding scheme of the resistive layer. For resistive MPGDs, this \textit{rate effect}, here referring to a device's ability to maintain its gain under a certain event rate per unit area, has been theoretically assessed using various methods. For resistive plane configurations like the one discussed here, previous work estimated the gain reduction as a function of the particle flux $\Phi$ [Hz cm$^{-2}$] for different surface resistivities \cite{GALAN2013229}. This estimation was achieved by taking the limit of the solution to the Telegraph equation \cite{Dixit2004} in the case of a current being imposed on an infinitely extending resistive layer. While this approach captured the qualitative behavior observed in experimental data, it does not account for the boundaries of the resistive readout.
Another approach, focusing on the µRWELL detector, was presented by G. Bencivenni et al. \cite{Bencivenni_2015}. They provided the average resistance to ground $\bar{\Omega}$ given the irradiated current over a circular area with a radius $r_b$ at the center of a circular resistive layer. This resistance could be expressed in terms of the layer's radius $c$ as $\bar{\Omega}(c) \simeq R (2c-r_b) /2\pi r_b$, where the voltage drop can then be calculated using Ohm's first law. A numerical approach was used in another study conducted by Z. Fang et al. \cite{FANG2022166615}. They represented the resistive layer as a discrete electrical system, thus enabling the assessment of the local voltage drops and rate effects for the more complex charge evacuation schemes found in various µRWELL prototypes \cite{Bencivenni2019} resulting from a position dependent irradiation current density. This work presents two alternative models, one giving an analytical expression for the potential drop on a finite square resistive layer, and a numerical method based on the Finite Element Method (FEM), where the Maxwell equations are solved under the quasi-static limit. The latter method provides the flexibility to handle systems with greater complexity that are technically more challenging to represent using a circuit-based approach, e.g., vertical charge evacuation through embedded resistors \cite{CHEFDEVILLE2016510,Alviggi_2017}, or multiple DLC layers \cite{Bencivenni2019,Iodice2020,Iodice2020b}.\\

First, the concept of the voltage drop across a resistive layer under an externally impressed direct current (DC) source will be reviewed, based on which a more complete description of the voltage drop across the finite resistive layer than those discussed above was derived. After this, the gain reduction will be estimated using numerical solutions of the potential in the $10 \times 10$~cm$^2$ multi-channel design of the resistive anode of Ref. \cite{lisowska2025photocathode} on which a current density from the particle flux is imparted. Using this, the maximum surface resistivity value that allows the resistive PICOSEC design to be effectively operated within a $\pi$ beam at the H4 Secondary Beam Line facility of the CERN Super Proton Synchrotron (SPS) will be estimated. 

\subsubsection{Voltage drop across a thin resistive layer}
\noindent In this part, the resistive configuration depicted in the inset of the top Fig. \ref{fig_concept} will be examined. This readout structure involves a thin resistive layer electrically connected to the ground and situated along its entire outer boundary at coordinates $x = 0, a$ and $y = 0, b$, which is positioned between the gas gap and the insulating layer. In addition, an amplification field in the gas gap of strength $E_a$ is applied. Given a charge $q$ deposited on the resistive layer at $t=0$, currents will start flowing in the resistive layer to compensate for this newly arrived charge, resulting in its seeming decay. As discussed in the work of W. Riegler \cite{Riegler2016}, the dynamics are governed by the infinite number of time constants, the largest of which when $a = b$ can be approximated to
\begin{equation}\label{eq: tau11}
\tau_{11} \approx \frac{1}{2 \pi^2} R\left(\varepsilon_1 \frac{a^2}{d}+\varepsilon_2 \frac{a^2}{g}\right)=\frac{1}{2 \pi^2} R\left(C_1+C_2\right) \, ,
\end{equation}
with $C_1$ and $C_2$ denoting the capacitance from the resistive layer to the readout and mesh, respectively, as if it were treated as a metallic layer. It dictates the time scale over which the charge is horizontally drained by the resistive layer to the ground frame. This happens at a rate set by this maximal time constant of the system, e.g., for $g = 128$ µm, $d = 60$ µm, $a = b = 10$ cm, $\varepsilon_1 = 3.7$, and $R = 10$ M$\Omega/\Box$ it yields $\tau_{11} = 3.12$ ms. As indicated by Eq. \eqref{eq: tau11}, when the endpoint of the resistive layer is shifted to a greater distance, this time constant increases.\\

Given several charges deposited in the same region in rapid succession, the combined effects of their electric fields within the amplification gap will lead to a localized reduction in $E_a$. In the continues limit, a constant DC-current can be applied to the resistive layer represented as $q(t)=I_0 t$, rather than introducing a single charge of $q(t)=q\Theta(t)$ \cite{LIPPMANN2006127}. The steady-state solution can then be found in the Laplace space through $\lim_{t\to \infty}\phi(\mathbf{x},t) = \lim_{s\to 0}s \phi(\mathbf{x}, s)$, resulting in the potential on the resistive layer to be
\begin{equation}\label{eq: Rate V drop on layer}
\begin{aligned}
\phi_\text{pt}(x, y) & =\frac{4 I_0 R}{a b} \sum_{l, m=1}^{\infty} \frac{\sin \left(\frac{\pi l x}{a}\right) \sin \left(\frac{\pi l x_0}{a}\right) \sin \left(\frac{\pi m y}{b}\right) \sin \left(\frac{\pi m y_0}{b}\right)}{\pi^2\left(\frac{l^2}{a^2}+\frac{m^2}{b^2}\right)} \\
& =2 I_0 R \sum_{l=1}^{\infty} \frac{\sinh \left(l \pi\left(a-x_{>}\right) / b\right) \sinh \left(l \pi x_{<} / b\right)}{\pi l \sinh (l \pi a / b)} \\ &~~~~~~~\times \sin \left(\frac{l \pi y_0}{b}\right) \sin \left(\frac{l \pi y}{b}\right)\, ,
\end{aligned}
\end{equation}
where the following notation is used
\begin{equation}
    x_> :=\begin{cases}
        x_0 \qquad x < x_0\\
        x~~ \qquad x \geq x_0
    \end{cases}\, , \qquad
    x_< :=\begin{cases}
        x~~ \qquad x < x_0\\
        x_0 \qquad x \geq x_0
    \end{cases}\, .
\end{equation}
In essence, this is the Ohmic relation between the voltage drop, resistance to ground and the injected current. It is worth noting that the dependence on the size and dielectric constants of the gas gap and insulating layer are absent in the steady-state solution, which is fully determined by the electrical properties and dimensions of the resistive layer. Eq. \eqref{eq: Rate V drop on layer} can be used the determine the potential drop over the resistive layer away form the beam spot. However, given the point-like injection of a current $I_0\delta(x-x_0)\delta(y-y_0)$ the solution diverges for $(x,y) \to (x_0,y_0)$. To account for a spatially extending beam profile, a uniform distribution of the events across a circular disk with radius $r_b$ is assumed. When centering the disk at a position $(x_c,y_c)$, the voltage drop across the resistive layer is given by

\begin{align} \label{eq: disk voltage drop}
\phi_{\text {disk }}(x, y)&=\frac{8 R I}{a b r_b} \sum_{l, m = 1}^\infty \frac{J_1\left(r_b k_{l m}\right)}{ k^3_{l m}} \sin \left(\frac{\pi l x}{a}\right) \sin \left(\frac{\pi m y}{b}\right)\nonumber\\&~~~~~~~~\times \sin \left(\frac{\pi l x_c}{a}\right) \sin \left(\frac{\pi m y_c}{b}\right)\, ,
\end{align}
where $I$ denotes the total current over the disk, $J_n(x)$ the n'th order Bessel function of the first kind, and 
\begin{equation}
k_{l m}:=\pi \sqrt{\frac{l^2}{a^2}+\frac{m^2}{b^2}}\, .
\end{equation} 
To evaluate Eq. \eqref{eq: disk voltage drop} numerically, an error term can be defined that quantifies the loss incurred by truncating the summation at maximum values $l \leq L_\text{max}$ and $m \leq M_\text{max}$, thereby neglecting the tail of the summation. Writing the error as $\epsilon = \| \text { tail } \|_{\infty}$, the evaluation can be performed up to
\begin{equation}
    L_\text{max} \approx \frac{a \kappa}{\pi} \qquad \text{and} \qquad M_\text{max} \approx \frac{b \kappa}{\pi} \, ,
\end{equation}
with 
\begin{equation}
\kappa=\left[\frac{32 \sqrt{2}}{3 \pi^{7 / 2}} \frac{R I}{r_b^{3 / 2} \varepsilon}\right]^{2 / 3} \, .
\end{equation}
Taking $\epsilon = 10$ mV for a 10 x 10 cm$^2$ resistive layer of 10 M$\Omega/\Box$ on which a constant current of $1$ µA is impressed uniformly on disk of radius 7.5 mm, the sum can be truncated to $L_\text{max} = M_\text{max} \approx 180$.
\\

\begin{figure}[t!]
\centering 
\includegraphics[width=.42\textwidth]{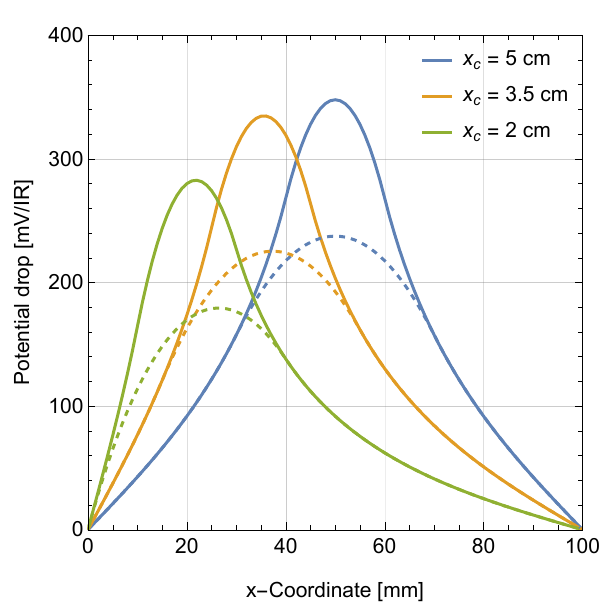}
\caption{\label{fig: voltage drop eq} Cross-section at $y = 50$ mm of the potential drop across a 10 x 10 cm$^2$ grounded resistive layer with surface resistivity R with an impressed total current $I$ across a disk centered at $(x_c,y_c = 5 \text{ cm})$ with radius $r = 1$ cm (full lines) and $r = 2$ cm (dashed lines).}
\end{figure}

An example is presented in Fig. \ref{fig: voltage drop eq} for a 10 x 10 cm$^2$ resistive layer irradiated with two different sized beam spots.
Given a central position of the beam spot, the voltage drop caused by the constant current deposition on the resistive readout structure is at its highest at the center of the beam profile on the active area. Subsequently, it gradually tapers as it extends towards the periphery of the beam spot, and its impacts can be detected throughout the entire resistive layer. If the total current $I$ is kept constant, the potential drops towards the Dirichlet walls is identical outside disks of different radii. However, the potential inside the disk tend to drop for larger $r_b$ as the current is divided over a larger areas. Furthermore, as the current is more efficiently drained the closer it is impressed to the edges of the resistive layer, the maximum voltage drop is not necessarily at the center of the disk when shifting the disk towards edge of the border, breaking the symmetry of the system. Using Eq. \eqref{eq: disk voltage drop}, the average potential across the disk is then given by
\begin{equation}\label{eq: disk phi av}
\bar{\phi}_\text{disk}\left(x_c, y_c\right)=\frac{16 R I}{a b r_b^2} \sum_{l, m = 1}^\infty \frac{J_1^2\left(r_b k_{lm}\right) }{k_{lm}^4}\sin ^2\left(\frac{\pi l x_c}{a}\right) \sin ^2\left(\frac{\pi m y_c}{b}\right)\, ,
\end{equation}
which is dependent on the location of the beam spot on the resistive layer and the constant current $I$. As will be explored in the next sub-section, when the potential on the resistive layer changes, the gain will be affected, modifying the value of $I(\bar{\phi}_\text{disk}) \propto \exp(c_1 \bar{\phi}_\text{disk})$, with proportionality constant $c_1$. This in turn, introduces a feedback process, that will influence $\bar{\phi}_\text{disk}$. As a result, the equilibrium value of $\bar{\phi}_\text{disk}$ of Eq. \eqref{eq: disk phi av} can be expressed in terms of a Lambert W function \cite{LambertW}.
\subsubsection{Estimating gain reduction at high particle fluxes}\label{sec: gain reduction}
\begin{figure}[t!]
\centering 
\includegraphics[height=.4\textwidth]{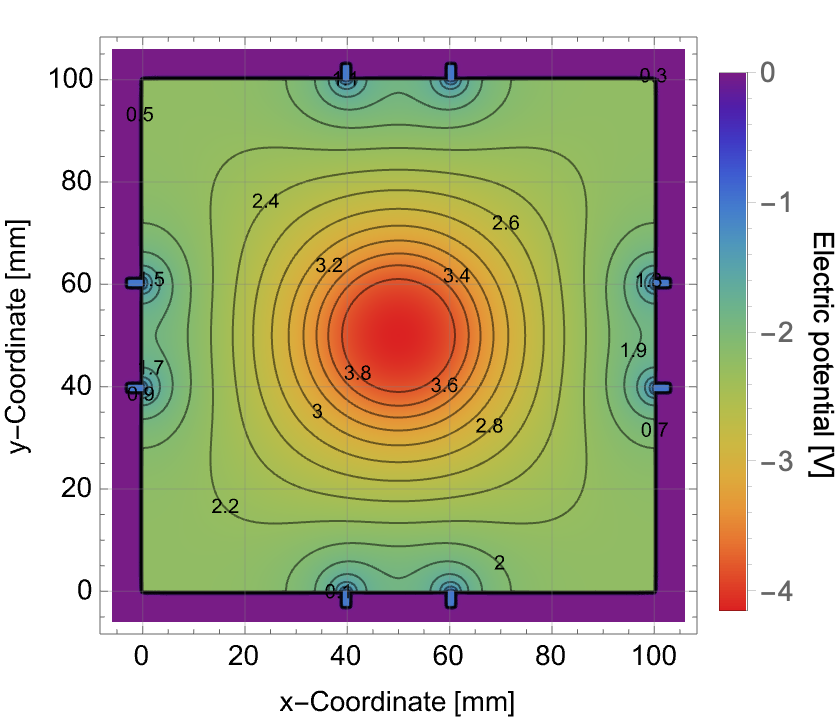}
\caption{\label{fig: PICOSEC rate drop} Electric potential $\phi(x,y)$ across the resistive layer due to the injection of a current density coming from a $15$ mm wide circular beam with a flux of $\Phi = 5.66\cdot10^5$ cm$^{-2}$s$^{-1}$ and effective detector gain $G_e = 10^6$.}
\end{figure}

\noindent Due to spatial constraints on the Printed Circuit Board (PCB), the final design of the resistive anode layer uses eight termination points as shown in Fig. \ref{fig: PICOSEC rate drop}, instead of connecting it along its entire outer edge as was the premise before. A $10$ M$\Omega$ resistor is connected to each of these points to ensure the minimal impedance at every point on the active area to avoid destructive capabilities of the discharges resulting in the damaging of the DLC layer. These additions require the use of a FEM approach to determine the steady-state solution of the voltage drop across the resistive layer given a circular $\pi$ beam profile with a rate of $1.9$ MHz. For the exponential relation between the amplification gain and the voltage difference of the mesh and anode, the resulting decrease in gain can be estimated. \\

Given the $10 \times 10$ cm$^2$ geometry shown in Fig. \eqref{fig: PICOSEC rate drop}, a constant boundary DC-current source $\mathbf{j}_e(x,y,t)$ is empressed on the resistive layer starting at $t = 0$. As an ansatz, the beam spot is taken to be circular with a radius of $r_b$ centered in the active area. In polar coordinates, the externally applied current density can be expressed in terms of the particle flux per surface area $\Phi$ as
\begin{equation}\label{eq: Rate impressed je}
    \mathbf{j}_e(r,t) =-e_0 n_p G_e\Phi\Theta(r_b-r-a)\Theta(t)\hat{\mathbf{z}} \, ,
\end{equation}
where  $e_0$ denotes the electron charge, and $G_e$ is the effective detector gain given an applied voltage $V_a$ on the anode. The number of primary charges $n_p$ depends on the quantum efficiency of the photocathode used, and for what follows will be taken as $n_p = 5$, which is a conservative overestimation for a DLC photocathode \cite{Sohl2020}. In addition, the beam's radius is fixed to $r_b = 7.5$ mm. Solving Maxwell's field equations using the electrical currents module within the COMSOL\textsuperscript{\tiny\circledR} toolkit \cite{COMSOL}, the steady-state solution was obtained for various values of $G_e$, $\Phi$, and $R$. An example of the potential value across the resistive layer is shown in Fig. \ref{fig: PICOSEC rate drop}. Unlike Fig. \ref{fig: voltage drop eq}, the potential does not reach zero at the connection points of the resistive layer due to the presence of the termination resistors. Following Ohm's law, this sets a lower bound to the rate capabilities even if $R\to 0$, since a global voltage drop will occur over the entire layer due to the constant current flowing through the 1.25 M$\Omega$ equivalent resistor. What is more, due to the symmetry of the system, the current is equally drained over all connection points, while shifting the position of the beam towards the corners of the active area, an increase in the voltage drop was found proportional to the increased current that flowing over the two neighboring resistors. \\

\begin{figure}[t!]
\centering 
\includegraphics[height=.4\textwidth]{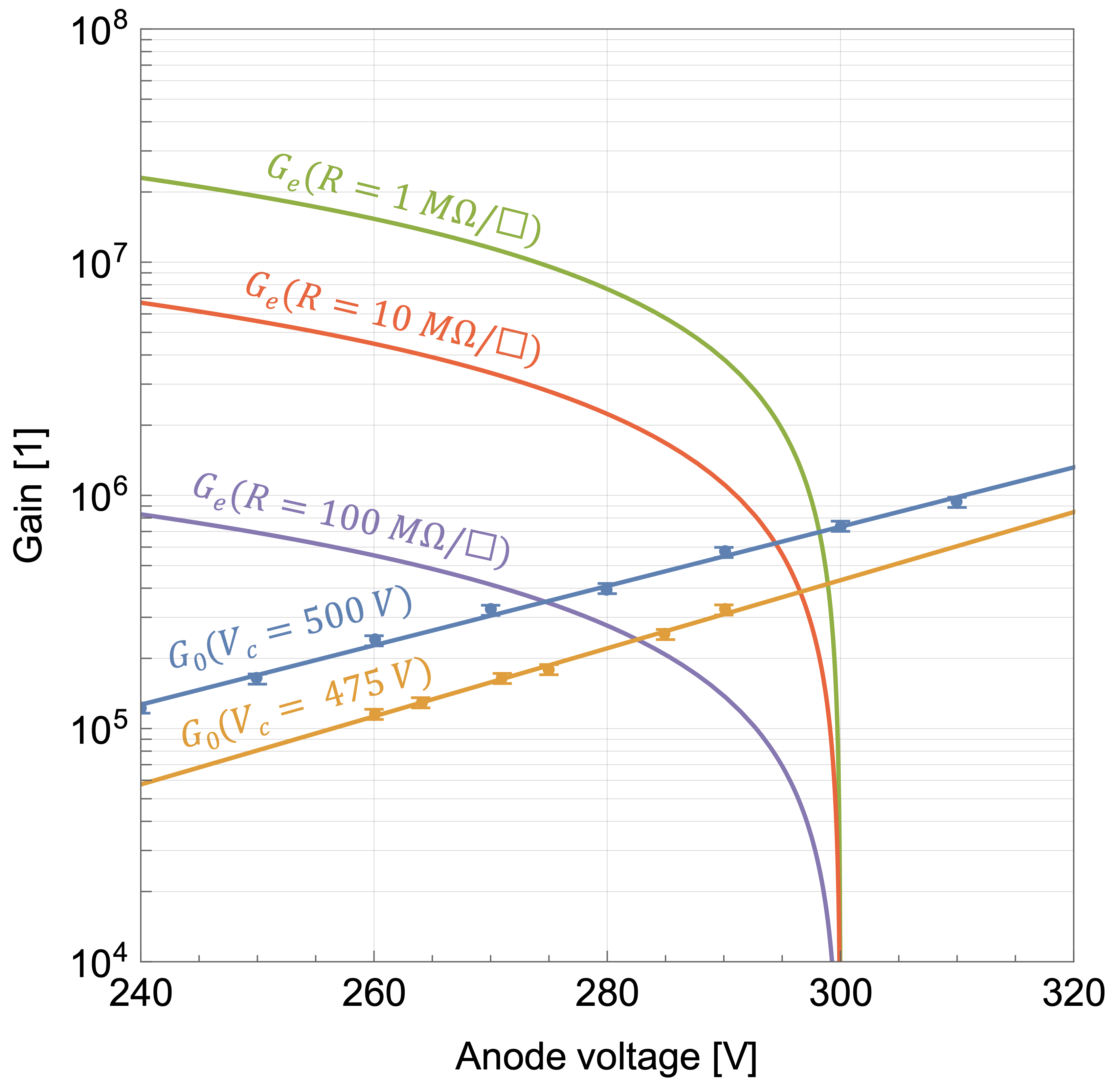}
\caption{\label{fig: PICOSEC voltage drop} Computed relation between the gain $G_e$ of the externally applied current density $j_e$ on the resistive layer as a function of the average voltage drop $V_a + \bar{\phi}$ on the anode. Furthermore, the measured gain is depicted, represented as $G_0$, for diverse (pre-)amplifications, indicating the data points with markers, where a 5\% error is assumed. The full lines signify the exponential fits for the two gain curves, employing Eq. \eqref{eq: G vs Va}. The equilibrium state can be identified as the point of intersection between $G_0$ and $G_e$.}
\end{figure}

Given an amplification field resulting from the potential difference $V_a$ between the grounded mesh and the anode plane, an additional average voltage drop in the beam spot $\bar{\phi}:=\int_A \phi(x,y)\,dA/\pi r_b^2$ can be added. The result in the irradiated circular area for fixed $\Phi$, and $V_a$ is shown in Fig. \ref{fig: PICOSEC voltage drop}, where $G_e$ is plotted as a function of the average voltage in the beam spot $V_a+\bar{\phi}$ that it causes. Until now, the value $G_e$ was taken to be independent of the local alteration of the anode voltage due to the charge deposition on the resistive layer, while in actuality, it will be coupled
\begin{equation}\label{eq: G vs Va}
G=\exp\left[c_1 (V_a+\bar{\phi})+c_2\right] \, ,
\end{equation}
where the parameters $c_1$ and $c_2$ depend on the gas properties and detector field configuration. As sufficient time passes, an equilibrium state $G$ is established where the surface current density impressed on the resistive layer leads to a voltage reduction in the irradiated area, ultimately resulting in an effective gain that matches the generated surface current density as given by Eq. \eqref{eq: Rate impressed je}. To find this stable point, the free parameters in the above relation between the effective gain $G_0$ and $V_a$, i.e., the \textit{gain curve} of the detector, must be first be estimated in the absence of any rate effects. \\

The gain curve of the amplification region for the detector can be either obtained through simulation or experiment by varying the amplification field and counting the arrived charge per event. Going with the experimental approach, a resistive single-channel prototype was used, featuring a pre-amplification gap size of 200 µm. The current $I_a$ on the anode plane resulting from a uniform and  $\mathcal{O}(1)$ kHz low-intensity UV LED irradiation was measured using a pico-ampere meter. Knowing the single photoelectron event rate $f_{pe}$ through the areas of the pulse hight spectra for a fixed exposure time using a digital multichannel analyzer (MCA)\footnote{Amptek MCA-8000D, \href{https://www.amptek.com/products/multichannel-analyzers/mca-8000d-digital-multichannel-analyzer}{https://www.amptek.com/products/multichannel-analyzers/mca-8000d-digital-multichannel-analyzer.}}, the gain could be estimated using $G_0 = f_{pe}/e_0I_a$ for different (pre-)amplification fields. The results for two different cathode voltages are plotted in Fig. \ref{fig: PICOSEC voltage drop} alongside the fitted curves of Eq. \eqref{eq: G vs Va} to estimate the trends. Here, the anticipated exponential relationship can be observed, where the fit adequately describes the data.\\

\begin{figure*}[t!]
    \centering
    \includegraphics[height=0.39\textwidth]{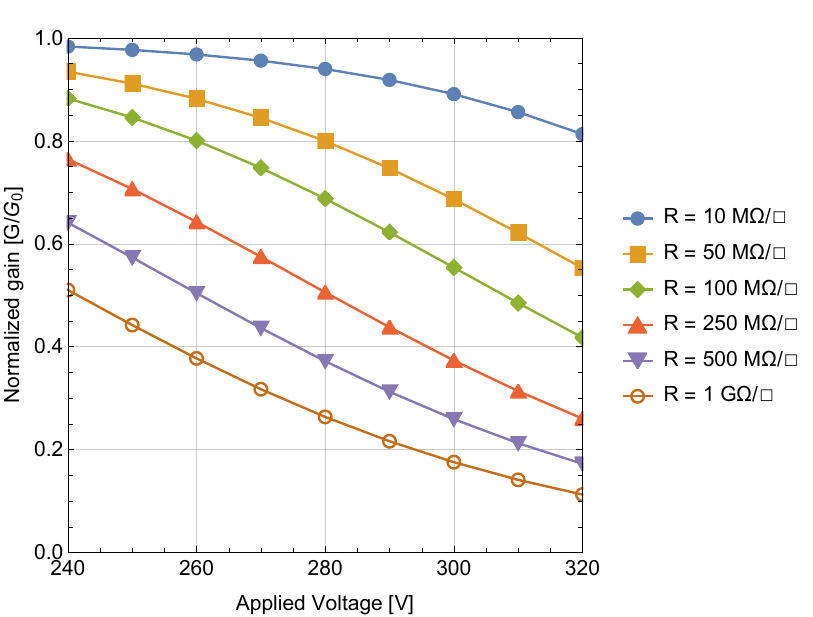}
    \includegraphics[height=0.39\textwidth]{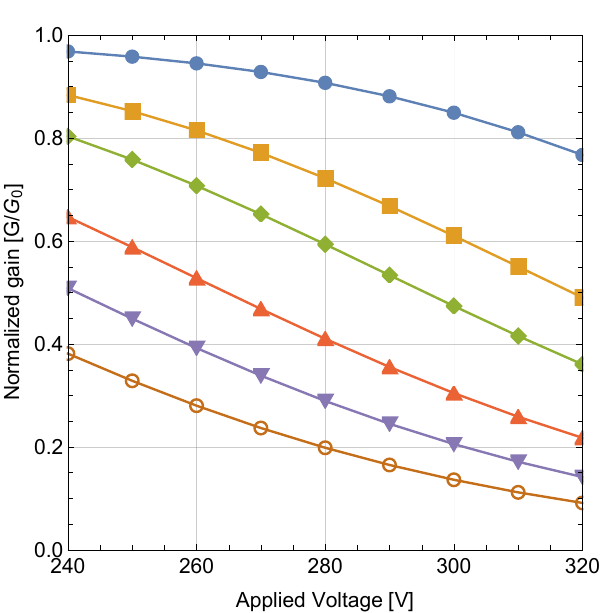}
    \caption{\label{fig:PICOSEC_gain_drop}
        Calculated normalized gain for the resistive PICOSEC as a function of the applied anode voltage $V_a$ for different surface resistivities and cathode voltage of $-475$ V (left panel) and $-500$ V (right panel).}
\end{figure*}

Following Eq. \eqref{eq: Rate impressed je} when $G_e = G_0$, $j_e$ constitutes the expected current density after irradiating over a long time scale, i.e., when the system is in equilibrium. Consequently, the steady-state solution for voltage drop and gain can be estimated by identifying the point of intersection between the two curves. In Fig. \ref{fig:PICOSEC_gain_drop} the gain drop curve is shown as a function of the anode voltage given a 1.9 MHz event rate. For operational voltages of $V_c = -475$ V and $V_a = 275$ V the expected gain is reduced by $5.24$\% and $8.87$\% for $R = 10$ M$\Omega/\Box$ and $R = 20$ M$\Omega/\Box$, respectively. While in this model, an ideal beam of pions is considered, in actuality, there will be contamination with particle showers from the interaction of the meson with the material of the experimental setup. This results in a significant increased energy deposit inside the detector geometry, yielding a temporary higher current density on the resistive layer, the degree of which is a function of the rate at which these showers occur. Hence, the results presented here can be regarded as the most favorable scenario. 

\subsubsection{Choice of surface resistivity} 
\noindent The steady-state solution of the local anode potential for different surface resistivities was obtained by combining the FEM numerical results of the simulated potential drop on the anode layer and the measured gain curve using low rate single photoelectron events. Considering the spatial variation of the voltage drop, denoted as $\phi(x, y)$, across the readout plane, implementing gain compensation via an increase in the applied anode voltage will uniformly elevate the amplification field. This elevation can lead to undesirably high gains in regions beyond the beam profile. Furthermore, in cases of non-constant event rates at intermittent intervals without incoming particles, the system will relax to its initial state, leading to an amplification field that surpasses the desired level. Notably, a below $20$\% gain drop is found for a surface resistivity of $20$ M$\Omega/\Box$.\\

Depending on the final HEP application requirements, this calculation can be repeated with the beam-relevant parameters. This outcome was taken into account, along with the minimal discharge protection prerequisites and a negligible reduction to the rising edge of the electron peak due to the delayed component of the signal (as discussed later), and the final surface resistivity value for production was fixed at $20$ M$\Omega/\Box$.\\

This simple resistive layout only allows the evacuation of the deposited charge along the edges of the geometry. Therefore, this design is not scalable for high-rate applications requiring extensive surface area coverage. Within the development of resistive layer-based readout structures within the MPGD family, more sophisticated `local' charge evacuation schemes have been devised that are suitable for large area coverage \cite{Bencivenni2019}. One such structure is realized through the implementation of draining vias in a double resistive layer layout, with the bottom layer ensuring minimal protective impedance \cite{CHEFDEVILLE2016510,Alviggi_2017,Bencivenni2019,Iodice2020,Iodice2020b}. These types of rapid grounding techniques hold the potential to overcome the typical rate limitations associated with large-area resistive readout structures and are now being studied for the multi-channel resistive PICOSEC designs \cite{Janssens2024}. 
\subsection{Signal integrity}
\noindent Incorporating a resistive layer into the readout structure of the robust PICOSEC MM design introduces an impact on the overall signal shape coming from the resistive layer's finite conductivity and the insulating laminate's presence. Crucial for timing measurements, this influence can adversely affect both the shape and amplitude of the leading-edge of the signal induced on the pad electrodes. The subject of the `transparency' of the resistive layer to a signal generated by charges moving above it and the induction on the readout electrodes located below has been a longstanding topic in detector R\&D \cite{BATTISTONI1982459}. A classic example is the contribution to the timing performance from the graphite layers, which form an integral part of contemporary (M)RPCs designed for the HV application and avalanche charge dissipation. Due to the finite conductivity of these layers, the reaction of this resistive element will result in the emergence of a delayed component to the signal, which negatively contributes to the signal amplitude due to its opposite polarity to the prompt current. It is now well understood that this contribution is negligible for the typical value of $\mathcal{O}(300)$ k$\Omega/\Box$ of this material. A seminal study on this topic was performed by G. Battistoni et al. in 1982  \cite{BATTISTONI1982459} who utilized a lumped element equivalent circuit to estimate the signal transparency of a thin resistive cathode layer, which divides the gas volume of a MWPC from an insulating layer containing AC-coupled closely spaced strip electrodes. As described in Ref. \cite{Riegler2016}, this approach hinges on the ability to define a capacitances between the resistive layer and electrodes, which only holds at late times. Instead, to be compatible with the quasi-static Maxwell equations, an extension to the the Ramo-Shockley theorem \cite{Shockley1938,Ramo1939} can be used to estimate the contribution from the resistive material to the induction of the electron peak found in the resistive PICOSEC design.
\subsubsection{Time-dependent weighting potentials}
\noindent The induction of signal on electrodes in the presence of resistive materials can be described by using an extended form of the Ramo-Shockley theorem for conductive media. The contribution of the material resistivity to the signal formation is included in the time evolution of the weighting potential, which is the solution of the Maxwell equations in the quasi-static limit \cite{Gatti1982,Riegler2002,Riegler2004,Riegler2019}. Consider a resistive detector with $N$ electrodes indexed $i\in\{1,2,\dots,N-1,N\}$ and a point charge carrier $q$ that follows a path $\mathbf{x}_q$ parametrized by time $t$. The induced signal on the electrodes is then given by
\begin{equation}\label{eq: induced current}
I_i(t)=-\frac{q}{V_w} \int_0^t d t^{\prime}\, \mathbf{H}_i\left[\mathbf{x}_q\left(t^{\prime}\right), t-t^{\prime}\right] \cdot \dot{\mathbf{x}}_q\left(t^{\prime}\right) \, ,
\end{equation}
where the dot notation refers to the derivative with respect to time and the weighting vector of electrode $i$ is a function of its dynamic weighting potential $\Psi_i$ through
\begin{equation}
    \mathbf{H}_i(\mathbf{x}, t):=-\nabla \frac{\partial \Psi_i(\mathbf{x}, t) \Theta(t)}{\partial t} \, .
\end{equation}
Here, the time-dependent weighting potential is obtained by removing the drifting charge carriers and applying the boundary conditions
\begin{equation}
\Psi_i(\mathbf{x},t)|_{\mathbf{x}\in S_j} = \delta_{i,j}V_w \Theta(t) \, ,
\end{equation}
with $S_j$ denoting the surface of electrode indexed $j$, $\delta_{i,j}$ the Kronecker delta function and $\Theta(t)$ the Heaviside distribution. The contribution from the direct induction of signal sourced by the movement of the charge carrier (called the \textit{prompt component}), and the part coming from the response of the resistive materials (\textit{delayed component}) can be disentangled on the weighting potential level by $\Psi_i(\mathbf{x},t):= \psi^p_i(\mathbf{x}) + \psi^d_i(\mathbf{x},t)$, with the delayed weighting potential being defined as $\psi^d_i(\mathbf{x},0) = 0$.\\

In what follows the resistive geometry depicted in the top Fig. \ref{fig_concept} will be used, where a thin resistive layer with a surface resistivity $R$ positioned between the amplification gap of size $g$ and the insulating layer with thickness $d$ is terminated around its entire outer edge situated at $x = 0, 10$ cm and $y = 0, 10$ cm. Since the main focus is in the formation of the signal on the $1$ cm$^2$ square pad electrodes on the readout plane, only the movement of the charges below the micro-mesh will contribute to this. Therefore, the amplification gap structure in the top Fig. \ref{fig_concept} is approximated as a parallel plate geometry. While the analytical solution can be obtained using Ref. \cite{Riegler2016}, the evaluation of the non-closed solution converges slowly for the dimensions of the Micromegas. Instead, it was chosen to use the more computationally efficient  numerical FEM recipe found in \cite{Janssens2022,janssens2024resistive}. Given a Townsend avalanche that occurs in the amplification region, the contribution of the delayed component to the electron peak was assessed as a function of the parameter $R$ and the relative position of the avalanche with respect to the pad electrode. The dynamic weighting potential was calculated for the central pad, i.e., $x_p = y_p = 5$ cm, for different surface resistivities and subsequently integrated into a Monte Carlo model for the avalanche development.

\subsubsection{Charge carrier simulation}
\noindent In order to calculate the expected change of the leading-edge of the signal induced on the $1\times1$ cm$^2$ readout pads with the introduction of the resistive layout, the above formalism is used in the open-source Monte Carlo simulation toolkit Garfield++ \cite{Garfield}. With the dynamic weighting potential being obtained numerically, it can be convoluted with the simulated drift lines of the electron and ions inside the gas \cite{janssens_2024_ef88z-72q34}. \\

For the propagation of these charge carriers through the gas, the applied electric field configuration needs to be defined. Similarly to the approach used to obtain the dynamic weighting potential, a FEM approach is used to solve the Laplace equation. Using unit cell a of the geometry, mirror periodicity is used on the xz- and yz-boundaries, while Dirichlet boundary conditions are imposed for the cathode (-455 V), micro-mesh (0 V), and anode (275 V). Having the electric field map, the COMPASS gas mixture with a Penning transfer rate of $r_p = 0.5$ \cite{BORTFELDT2021165049} is used to microscopically simulate the movement of the electrons and  the formation of a Townsend avalanche starting from a single photoelectron \cite{Schindler2012}. For the ions, measured reduced mobilities of Ne$^+$ in pure Ne \cite{hornbeck1951drift, Beaty1961} were used to perform a Monte Carlo integration of their drift paths, neglecting any possible ion chemistry inside the gas \cite{Kalkan_2015,Marques:2019jwi}. As the electrons traverse the micro-mesh into the amplification region, the signal induction process on the readout pads starts. The resulting induced current can be treated as an ideal current source in an LTSpice model \cite{LTSpice} that incorporates of the pad capacitance along with the current amplifier's response.
\subsubsection{Electron-peak amplitude}
\begin{figure}[t!]
\centering 
\includegraphics[height=.42\textwidth]{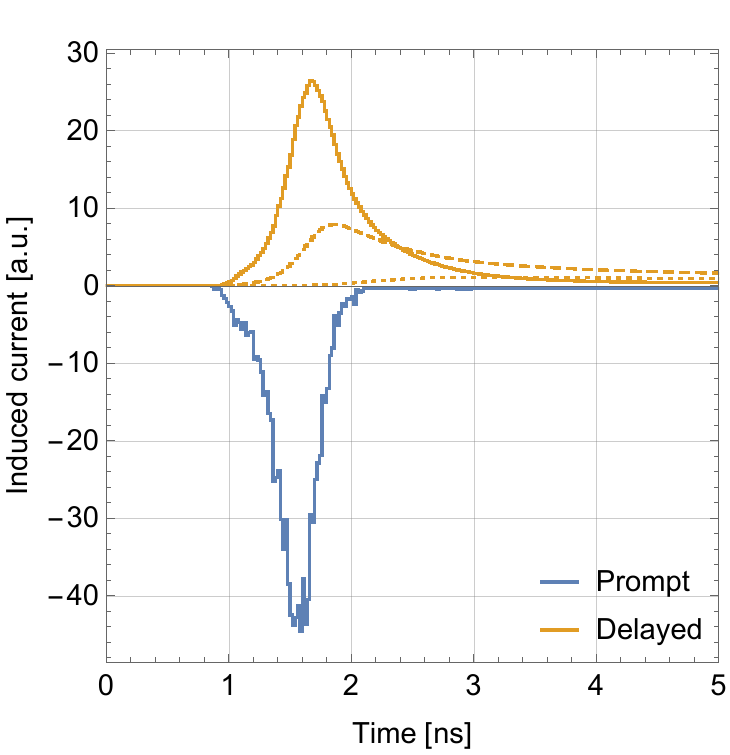}
\includegraphics[height=.42\textwidth]{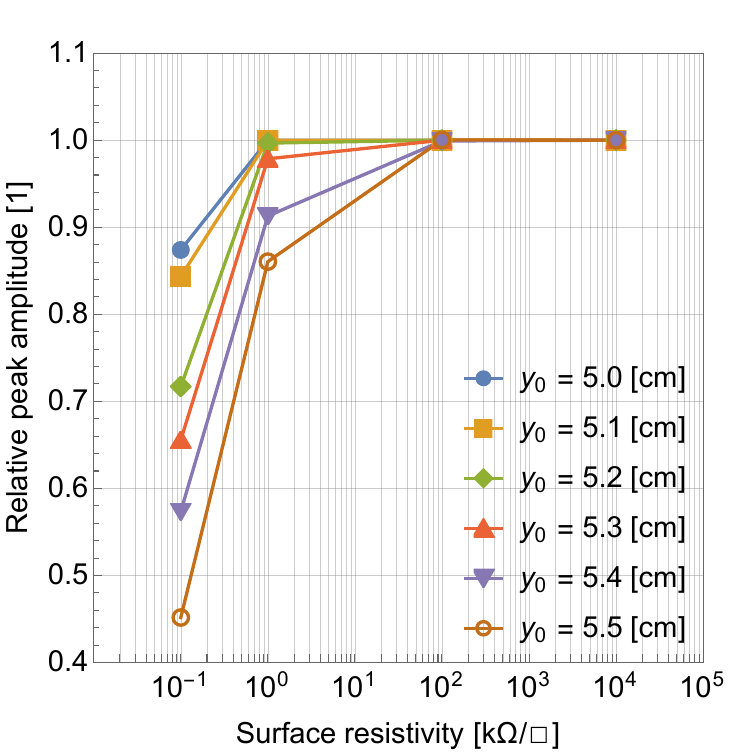}
\caption{\label{fig: PICOSEC amp drop} Top: The prompt and delayed components of the induced signal on the $1$ cm wide square pad, which is positioned at the center of the active area ($x_p=y_p=5$ cm), are determined for a Townsend avalanche initiated at the top of the amplification gap, specifically at the point centered at $x_0=0$ and $y_0=5.4$ cm. The result is given for $R = 100$ $\Omega/\Box$  (full line), $R = 1$ $k\Omega/\Box$ (dashed line), and $R = 100$ $k\Omega/\Box$ (dotted line). Bottom: The relative peak amplitude as a function of the surface resistivity for different avalanche positions on the pad where $x_0=0$. The plot markers represent the calculated values, while the lines are to guide the eye.}
\end{figure}

\noindent An example prompt signal $I^p$ and delayed signal $I^d$ from a Townsend avalanche is shown in Fig. \ref{fig: PICOSEC amp drop} (top) for three different surface resistivities. Although the ion tail is indeed accounted for, its amplitude is significantly smaller in comparison to the electron signal, causing the ion tail to be nearly imperceptible due to the fine signal binning. The amplitude of the prompt component resulting from the instantaneous induction of signal from the movement of both the electrons and ions below the mesh is subject to the relative permittivity $\varepsilon_r$ and thickness of the insulating layer. Compared to the non-resistive design where this laminate is absent, this results in a fundamental signal reduction by a factor $f_p$, which in the limit of $w_x\gg g$ can be written as 
\begin{equation}
    f_p :=\frac{E^p(d)}{ E^p(0)}= 1-\frac{d}{d+ \varepsilon_r g} \, ,
\label{eq_signal_reduction_factor}
\end{equation}
where the prompt weighting field of a readout plane bellow an insulating layer was used
\begin{equation}
    \mathbf{E}^p(d) = \frac{\varepsilon_rV_w}{d+ \varepsilon_r g}\hat{\mathbf{z}}
\end{equation}
with the relative permittivity $\varepsilon_r$, the thickness of the insulating layer $d$, and the size of the amplification gap $g$. In the case of the resistive PICOSEC MM, this corresponds to a value of $f_p = 0.89$. The response of the resistive layer, as captured by the delayed component of the signal, occurs over a longer time as $R$ grows, such that for large surface resistivities this contribution can be neglected inside the initial $\approx 1$ ns time range of the electron peak. \\

As an indicator for the contribution of the resistive layer to the electron peak amplitude, the \textit{relative peak amplitude} is defined as the ratio between the prompt signal and the total signal: $\text{Min}\left[I^p(t)+I^d(t)\right]/\text{Min}\left[I^p(t)\right]$. This is plotted as a function of different surface resistivities in Fig. \ref{fig: PICOSEC amp drop} (bottom). A heightened sensitivity to the resistive layer's behavior can be observed at the periphery of the pad electrode due to the more immediate signal spreads over the adjacent channels. As $R \gtrsim 100~\text{k}\Omega/\Box$, a regime is entered in which the electron peak of the induced signal is virtually unaffected by the delayed component of the signal. This outcome remains consistent even when accounting for the response of the pulse amplifier that reads out the channels. This was accomplished by utilizing an LTSpice amplifier model, treating the induced signals as ideal current sources.\\

In conclusion, due to the occurrence of avalanches in the amplification region over the several mm wide area covered by the radiator's Cherenkov cone, coupled with the lack of prior knowledge regarding the particle's hit position, it is imperative to reduce the impact of the DLC layer's response on the leading-edge of the signal across the entire surface of the readout pad. It was found that for surface resistivities exceeding 100 k$\Omega/\Box$, the electron peak is unaffected by the delayed component on the signal. However, an overall amplitude reduction of around 11\% must be contended with due to the presence of the $60~\mu$m thick insulating layer between the amplification gap and the readout plane. This model can be further developed in order to describe the relation between the temporal resolution for the case of surface resistivities below 100 k$\Omega/\Box$. Furthermore, it has the potential to quantify the extent of enhancement in spatial resolution attributable to signal propagation to neighboring channels by including the weighting potentials of the adjacent pads in the calculation.

\section{Single channel resistive PICOSEC MM design}

\noindent The single-channel resistive-anode PICOSEC MM detector is based on the metal-anode design reported in \cite{utrobicic2025single}, modified to incorporate a resistive readout structure.The mechanical layout and internal detector components, including the chamber body, PEEK insert, drift-gap spacers, and MgF$_2$ Cherenkov radiator with deposited photocathode, were kept identical in order to isolate the impact of the resistive anode on detector performance. Only the Micromegas and outer readout boards were redesigned to accommodate the resistive layer and its biasing scheme. \\

\begin{figure}[!t]
\begin{center}
\includegraphics[width=.42\textwidth]{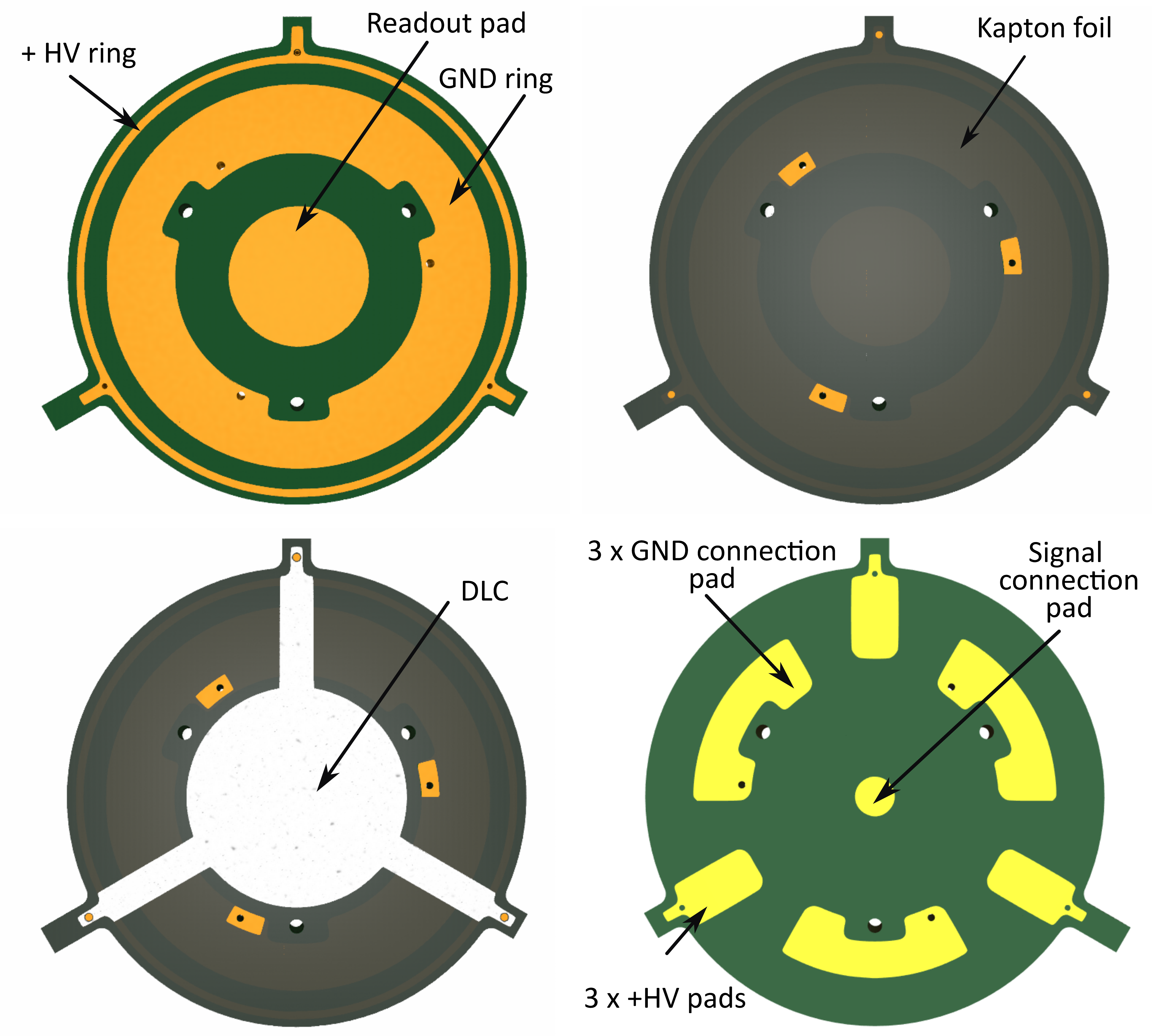}
\caption{Resistive Micromegas board top (upper left, upper right, and lower left) and bottom side layout (lower right). Upper left: Top copper layer that includes a central readout pad surrounded by the GND plane and a thin HV ring at the outer board perimeter. Upper right: Kapton foil covering the top copper surface. It includes three square openings that enable mesh connections to the GND, three holes to allow gas circulation, and three circular openings at the outer board edge for +HV connection. Bottom left: Circular DLC layer with three extensions toward the board edge for HV connection. Bottom right: Bottom copper layer that has a central signal connection pad, three GND connection pads and three HV connection pads.}
\label{fig_resistivePCB}
\end{center}
\end{figure}

The Micromegas board was manufactured from a double-layer FR4 substrate with a total thickness of 3.2 mm. The top and bottom layouts of the resistive board are shown in Fig. \ref{fig_resistivePCB}. The top side, shown in upper right Fig. \ref{fig_resistivePCB}, consists of a circular copper readout pad ($\varnothing$10 mm or $\varnothing$15 mm), surrounded by a grounded copper ring of 5~mm width, and an additional outer copper ring for the application of the positive high-voltage (HV) bias to the resistive layer. A $50~\mu$m thick Kapton foil carrying a patterned DLC resistive coating is s glued with 10~$\mu$m epoxy glue on top of the MM~board, as shown in upper right and bottom left Figs. \ref{fig_resistivePCB}. The resistive layer consists of a central active disk matching the readout pad diameter, along with three $\approx$~9~mm extensions reaching toward the edge of the MM PCB for +HV biasing.  The layout on the bottom side features a central pad for signal readout, three wider pads for the ground (GND) connection, and three shorter pads for the +HV connection, as illustrated in bottom right Fig. \ref{fig_resistivePCB}.\\\\

\begin{figure}[!t]
\begin{center}

\includegraphics[width=.42\textwidth]{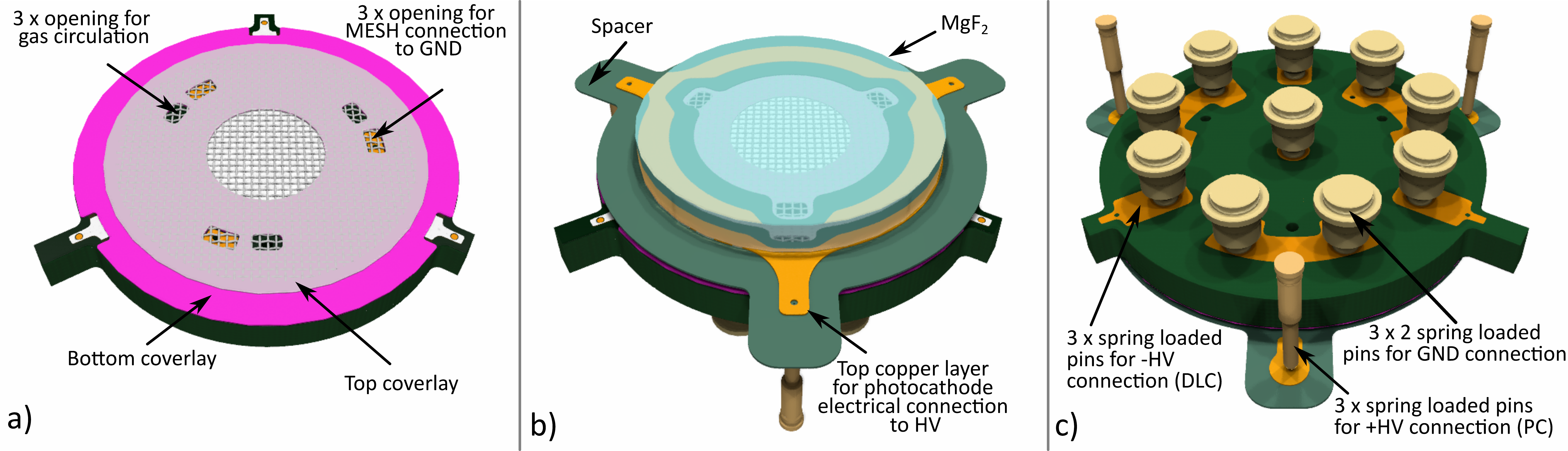}
\caption{Left: 3D model of a top side Resistive Micromegas board showing mesh integrated between the insulating layers. Middle: 3D model of Resistive MM board, spacer and MgF$_2$ crystal arrangement within the detector. Right: 3D model of spring-loaded pins arrangement for electrical connection of the mesh to the GND, photocathode to the -HV and DLC layer to the +HV.}
\label{fig_3D_model}
\end{center}
\end{figure}

The stainless steel micro-mesh, with a pitch of 45~$\mu$m and wire diameter of 18~$\mu$m, is embedded on top of the MM board between two insulating layers as shown in Fig. \ref{fig_3D_model} (left). Three pockets in both the top and bottom insulating layers provide the GND electrical contact to the mesh, while three additional apertures allow for gas circulation through the amplification region. As shown in the inset of the top Fig. \ref{fig_concept}, the thickness of the bottom insulating layer (nominally $\approx$128 µm) mechanically defines the amplification gap. The detector assembly is done in the same way as described in Ref. \cite{utrobicic2025single}, in which a spacer is placed above the top insulating layer to define the drift gap thickness and to provide the -HV electrical connection to the photocathode deposited on the bottom side of the MgF$_2$ crystal, as depicted in Fig. \ref{fig_3D_model} (middle). The spring-loaded pin system used for HV biasing and GND is shown in Fig.~\ref{fig_3D_model}  (left), ensures the electrical contact between the detector electrodes and the external circuitry. Compared with the metal anode detector, the resistive design has six ground connections instead of nine and distributes the resistive anode HV bias over three remaining contact points, each routed through an external 10 M$\Omega$  quenching resistor.  
Two types of single-channel resistive-anode prototype were produced at the CERN EP-DT-DD Micro-Pattern Technologies (MPT) workshop, featuring active areas of $\varnothing$10~mm and $\varnothing$15~mm. To estimate the fundamental signal attenuation factor in Eq. \ref{eq_signal_reduction_factor}, information on the insulating layer and the amplification gap thickness is required. Although metal and resistive anode detectors are nominally manufactured with identical bottom insulating layer thicknesses, small variations can arise from production tolerances. Therefore, a measurement of the distance between the anode surface and the top of the calendared micro-mesh was performed using a Hirox RH-2000 digital microscope \cite{HirroX} by focusing on surface imperfections on both electrodes through the mechanical movement of the apparatus. The measured amplification gap sizes are $(134.9 \pm 1.4)$~$\mu$m and $(139.4 \pm 1.4)$~$\mu$m for the metallic and resistive-anode detectors, respectively.

\section{Experimental verification}
\subsection{Measurements of the parasitic parameters of the resistive detector}

\begin{figure}[!t]
\begin{center}

\includegraphics[width=.42\textwidth]{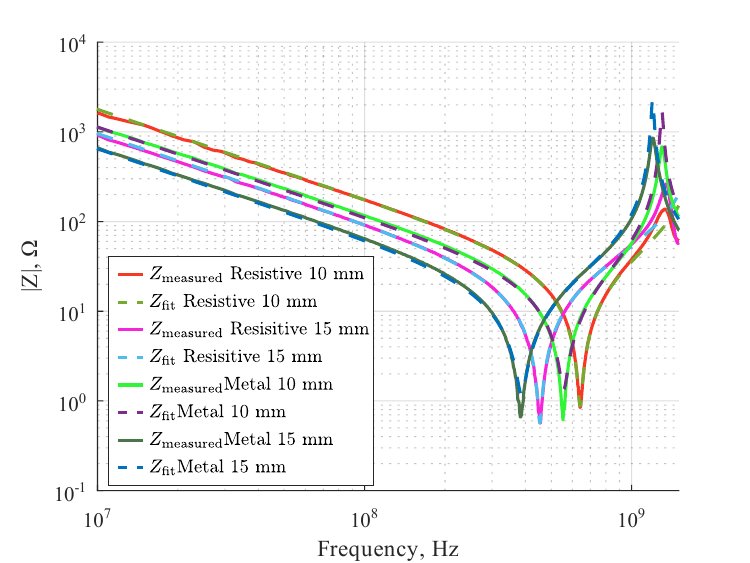}
\caption{Measured impedance of metal and resistive anode detectors with 10 mm and 15 mm diameter pad areas, with the fit overlayed as dash lines.}
\label{fig_impedance}
\end{center}
\end{figure}

\noindent The readout pad capacitance and signal path inductance are important parameters that strongly influence the shape of the measured electron signal \cite{utrobicic2025single}. The impedance of both the metal and resistive-anode ($\varnothing$10~mm and $\varnothing$15~mm) detectors was measured over a frequency range from 10~MHz to 1.5~GHz using a Siglent SVA1032X vector network analyzer (VNA). The measurement procedure and the subsequent impedance fitting method followed the same approach as described in Ref. \cite{utrobicic2025single}. Fig. \ref{fig_impedance} shows the measured impedance characteristics with the corresponding fits overlaid, and a summary of all fitted parameters is provided in Table \ref{tbl_fit_params}. \\

\begin{table}[!b]
\begin{center}
\caption{Parameters obtained by impedance fitting.}
\vspace{0.25 cm}
\small
\label{tbl_fit_params}
\begin{tabular}{l|l|l|l|l}

\multirow{2}{*}{\begin{tabular}[c]{@{}l@{}}Fit \\ parameter\end{tabular}} & 
\multirow{2}{*}{\begin{tabular}[c]{@{}l@{}}Met. anode\\  $\varnothing$10 mm\end{tabular}} & 
\multirow{2}{*}{\begin{tabular}[c]{@{}l@{}}Met. anode\\  $\varnothing$15 mm\end{tabular}} & 
\multirow{2}{*}{\begin{tabular}[c]{@{}l@{}}Res. anode\\  $\varnothing$10 mm\end{tabular}} & 
\multirow{2}{*}{\begin{tabular}[c]{@{}l@{}}Res. anode\\  $\varnothing$15 mm\end{tabular}} \\
            & & & & \\ \hline \hline
$C_{pad}$, pF    & 11.49$\pm$0.27 & 22.02$\pm$0.52 & 7.82$\pm$0.20 & 15.62$\pm$0.37 \\ \hline
$L_{\sigma}$, nH & 7.12$\pm$0.17  & 7.68$\pm$0.19  & 7.88$\pm$0.20  & 7.90$\pm$0.19 \\ \hline
$C_{con}$, pF    & 2.59$\pm$0.73  & 2.57$\pm$0.73  & 1.04$\pm$0.11  & 1.04$\pm$0.09 \\ \hline
$R_s$, $\Omega$  & 1.35$\pm$0.28  & 1.17$\pm$0.26  & 0.87$\pm$0.30  & 0.57$\pm$0.21 \\ 
\end{tabular}
\end{center}
\end{table}
It can be observed that the measured pad capacitance of the resistive-anode detector is $\approx$30\% lower than that of the metal-anode detector, while its parasitic inductance is slightly higher, possibly due to the reduced number of spring-loaded ground-connection pins. The signal connection capacitance C$_{con}$ is significantly lower in the resistive detector as it lacks the biasing trace and biasing resistor pad, which is implemented on the Outer Board of the metal anode detector. As expected, the resistive-anode detector exhibits the reduced C$_{pad}$ due to the addition of an insulating layer with resistive coating, but this reduction is larger than the theoretically predicted by relation Eq. \eqref{eq_capacitance}. This discrepancy may arise from the contribution of the parasitic capacitance, such as those introduced by the 200~$\mu$m bottom insulator covering the readout pad of the metal-anode detector, the capacitance between the readout pad and the surrounding ground ring, and the capacitance between the readout pad spring-loaded pin and the ground pins.

\subsection{Signal dynamics measurements}
\noindent To investigate the impact of an insulating layer with a resistive DLC coating added on top of the readout pad on the detector performance, measurements of the output signal dynamics were carried out for detectors with metal and resistive anodes using single photoelectron under laboratory conditions.
\begin{figure}[!t]
\begin{center}

\includegraphics[width=0.69\columnwidth]{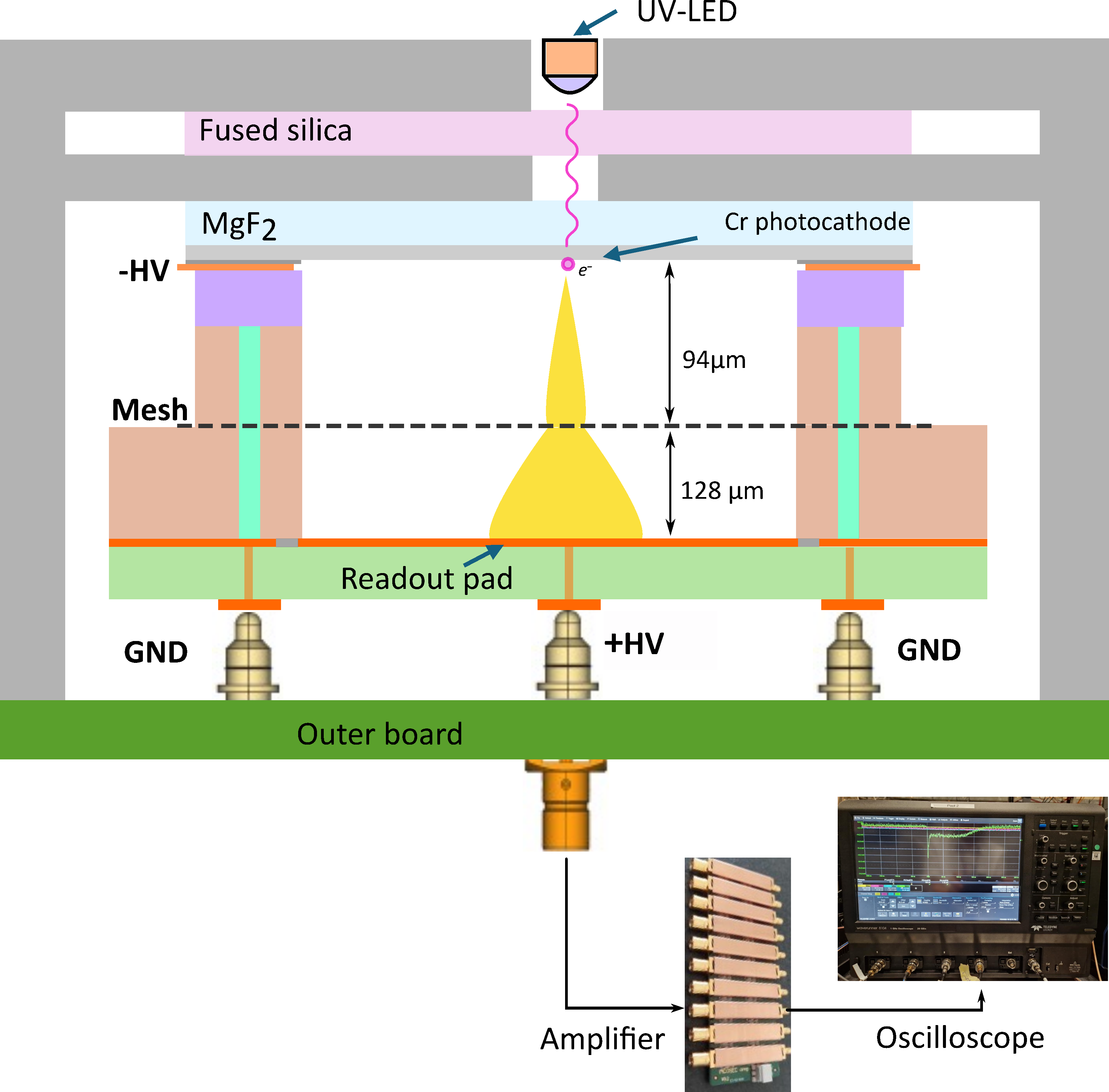}
\caption{Schematic representation of the experimental setup used for the single-photoelectron signal dynamics measurements. The PICOSEC MM detector is operated with a semi-transparent Cr photocathode on an MgF$_2$ radiator and illuminated by a pulsed UV LED. The induced signal on the readout pad is amplified by a custom preamplifier and digitized by a fast oscilloscope.}
\label{fig_single_pe}
\end{center}
\end{figure}
Measurements were performed on $\varnothing$10~mm active area metal or resistive anode detectors with a semi-transparent Cr photocathode.The amplification and drift gap thicknesses were 128~$\mu$m and 94~$\mu$m, respectively. The induced signal from the readout pad is amplified with a custom-made low-noise preamplifier and digitized at a sampling rate of 10~GS/s by a LeCroy WR8104 oscilloscope, as shown in Fig. \ref{fig_single_pe} \cite{hoarau2021rf,kovacic2022amp}.

\subsubsection{Waveform and PSD analysis for different field configuration}
\noindent Multiple runs were recorded at different anode or cathode voltages to analyze the shape of the signal. To obtain representative waveforms and reduce the noise in the ion tail region, the signals of the detector were averaged in both the frequency and time domain for the same selected e-peak amplitude range. In this way, it was possible to compare signal characteristics under different field settings.\\

\begin{figure}[!t]
\begin{center}

\includegraphics[width=0.7\columnwidth]{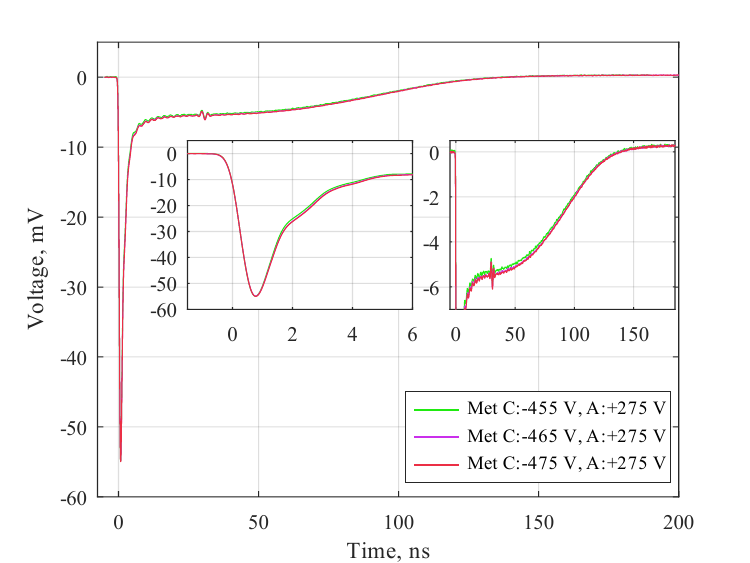}

\vspace{1mm}

\includegraphics[width=0.69\columnwidth]{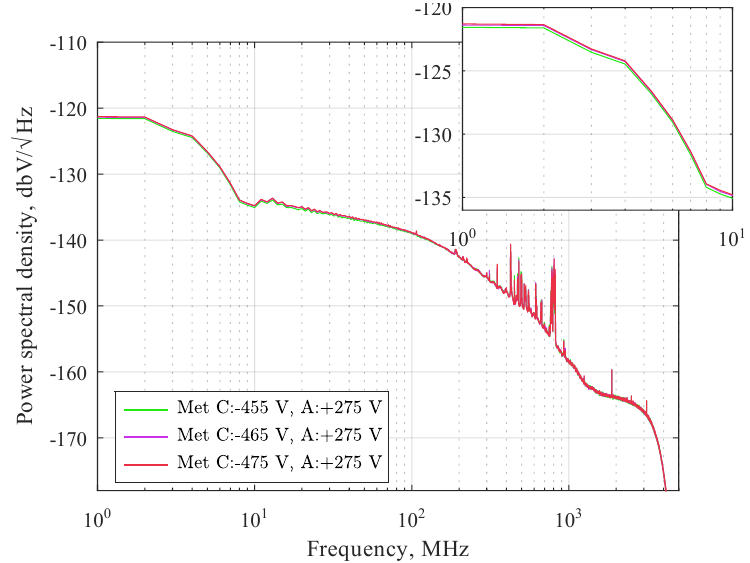}
\caption{Averaged waveforms (top) and power spectral density (bottom) for measurements with a metal detector at a fixed anode voltage of 275~V and cathode voltages of -455~V, -465~V, and -475~V.}
\label{fig_MET_fixed_anode_wf}
\end{center}
\end{figure}

\begin{figure}[!b]
\begin{center}

\includegraphics[width=0.7\columnwidth]{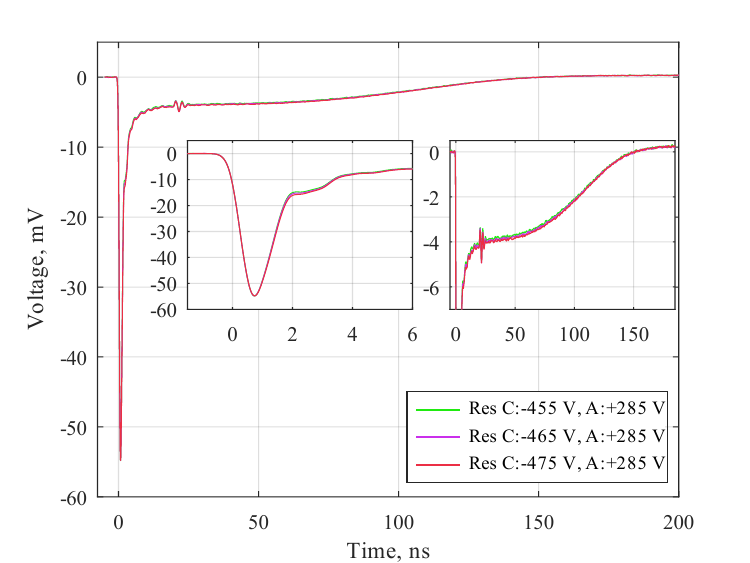}

\vspace{1mm}

\includegraphics[width=0.7\columnwidth]{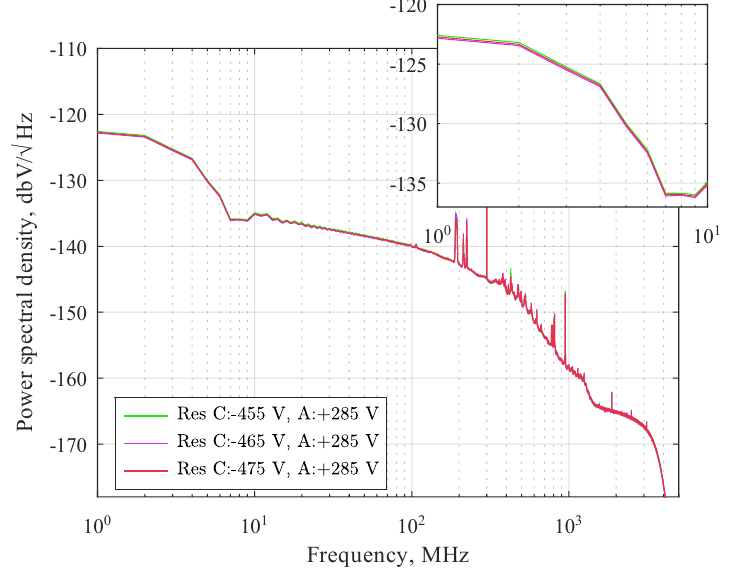}

\caption{Averaged waveforms (top) and power spectral density (bottom) for measurements with a resistive detector at a fixed anode voltage of 285~V and cathode voltages of -455~V, -465~V, and -475~V.}
\label{fig_RES_fixed_anode_wf}
\end{center}
\end{figure}

Fig. \ref{fig_MET_fixed_anode_wf}  and \ref{fig_RES_fixed_anode_wf} show the waveforms (top) and power spectral densities (bottom) obtained by averaging approx. 1000 waveforms or spectra of individual events recorded at a fixed anode voltage and three different cathode voltages for the metallic and resistive detectors. Signals with an e-peak amplitude between 55~mV and 60~mV were selected for analysis. It can be observed that the signal shape remains unchanged for different cathode voltage settings in both detector types, indicating that the signal shape is independent of the cathode voltage, i.e., the drift field strength.\\

\begin{figure}[!t]
\begin{center}

\includegraphics[width=0.7\columnwidth]{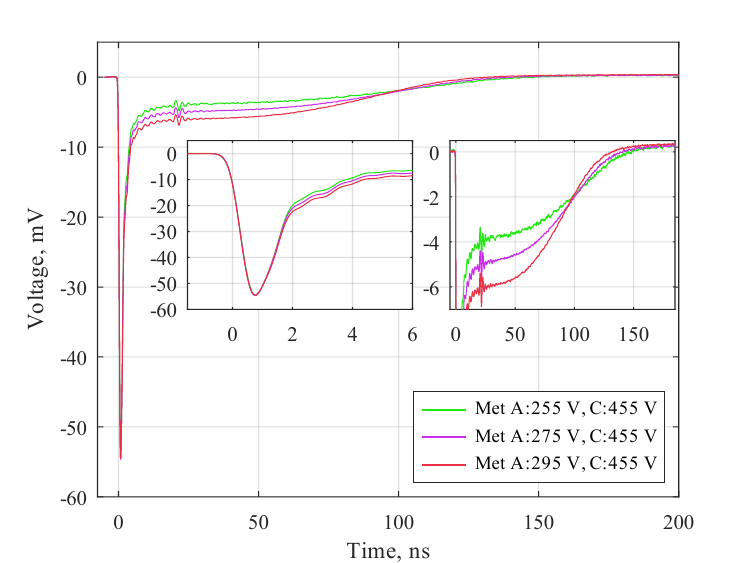}

\vspace{1mm}

\includegraphics[width=0.7\columnwidth]{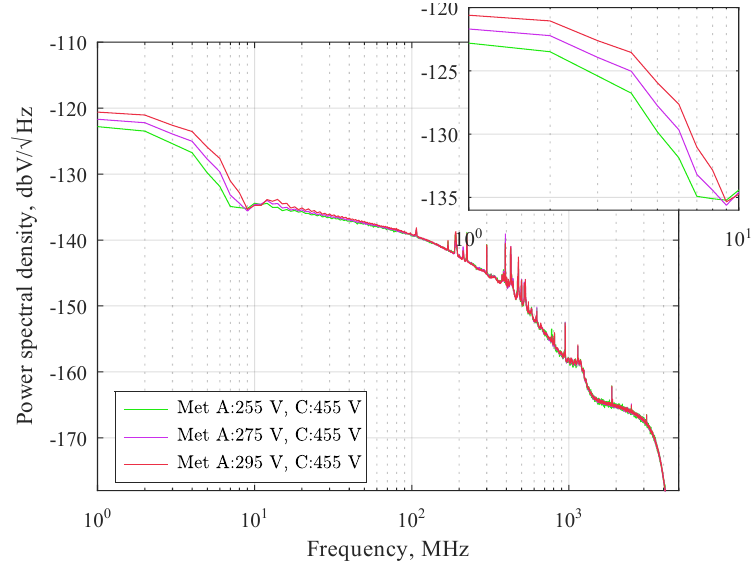}

\caption{Averaged waveforms (top) and power spectral density (bottom) for measurements with a metal detector at a fixed cathode voltage of -455~V and anode voltages of 255~V, 275~V, and 295~V.}
\label{fig_MET_fixed_cathode_wf}
\end{center}
\end{figure}

\begin{figure}[!b]
\begin{center}

\includegraphics[width=0.7\columnwidth]{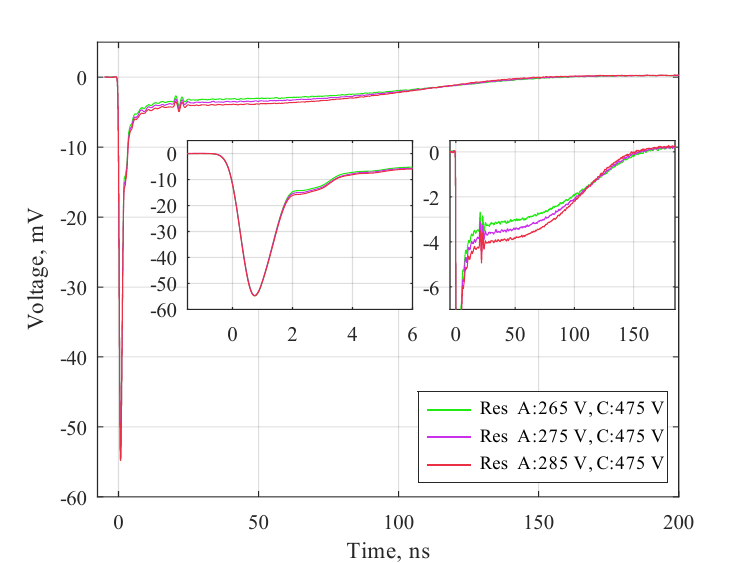}

\vspace{1mm}

\includegraphics[width=0.7\columnwidth]{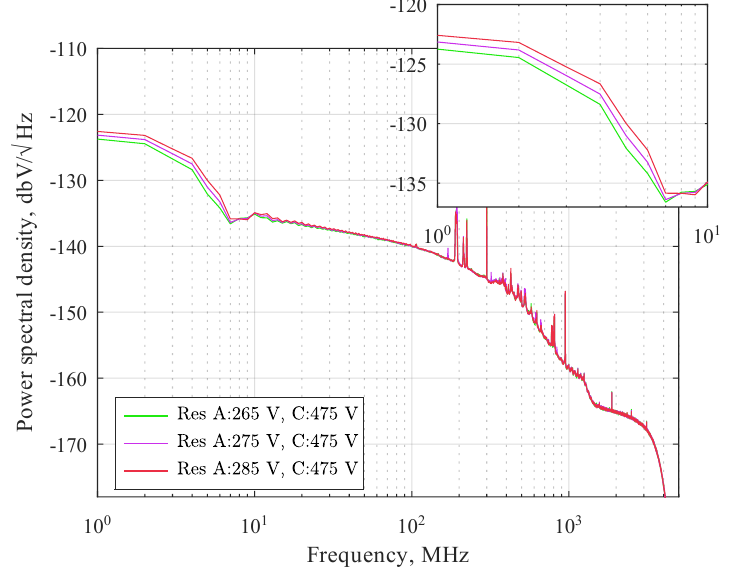}

\caption{Averaged 1000 waveforms (top) and power spectral density (bottom) for measurements with a resistive detector at a fixed cathode voltage of -475~V and anode voltages of 265~V, 275~V, and 285~V.}
\label{fig_RES_fixed_cathode_wf}
\end{center}
\end{figure}

The same analysis was performed for a fixed cathode voltage and three different anode voltages for both the metallic and resistive detectors. In the top and bottom Fig. \ref{fig_MET_fixed_cathode_wf} and \ref{fig_RES_fixed_cathode_wf}, it can be observed that only the shape of the ion tail is affected by the change of anode voltage, i.e., the amplification field, while the shape of the e-peak leading edge remains unchanged. This indicates that the dynamics of the measured e-peak signal are dominated by parasitic elements and the amplifier’s bandwidth, suggesting that the actual induced signal on the anode is significantly faster.






\subsubsection{Comparison of the detectors}
\begin{figure}[!t]
\begin{center}
\includegraphics[width=.99\columnwidth]{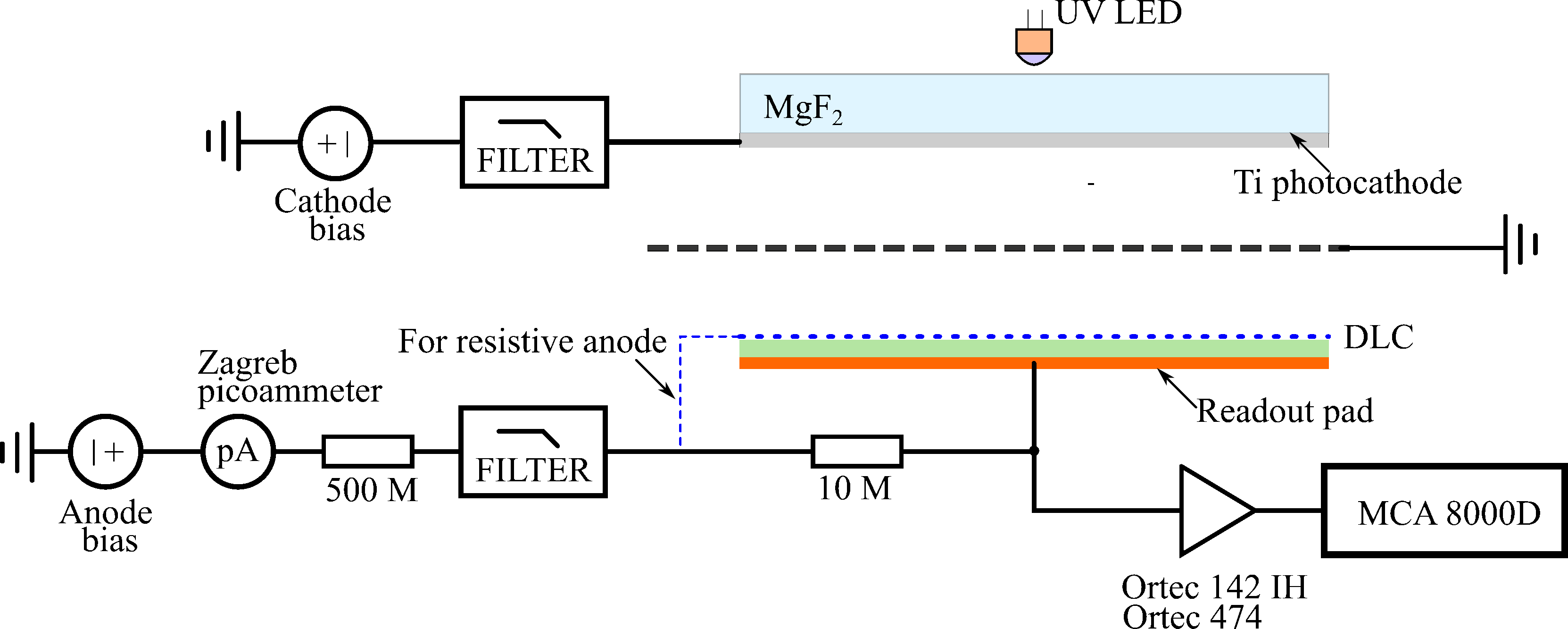}
\caption{Schematic of PICOSEC MM detector cross section and powering circuit.}
\label{fig_gain_measurements}
\end{center}
\end{figure}

Measurements for detectors with metal and resistive-anodes were performed at the same amplification in order to enable a direct comparison. Owing to small geometrical differences between the two detectors, the detector biasing were adjusted accordingly to achieve the same effective gain. As in Sec. \ref{sec: gain reduction}, the gain measurements were performed using a floating picoammeter~\cite{utrobicic2015floating} connected to the anode bias line. The readout electrode was connected to a charge preamplifier (ORTEC 142 IH) and a shaper (ORTEC 474), and finally to a digital MCA that recorded the event rate and amplitude spectrum, as shown in Fig. \ref{fig_gain_measurements}.\\

\begin{figure}[!t]
\begin{center}
\includegraphics[width=0.7\columnwidth]{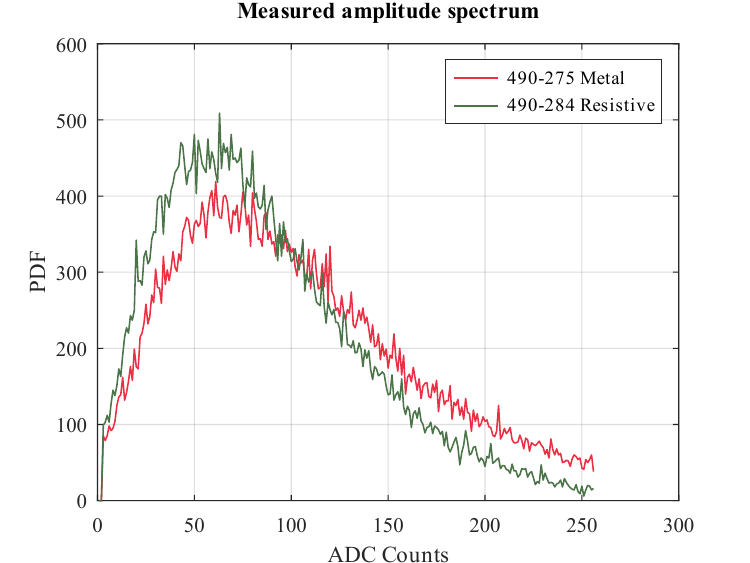}

\vspace{1 mm}

\includegraphics[width=0.7\columnwidth]{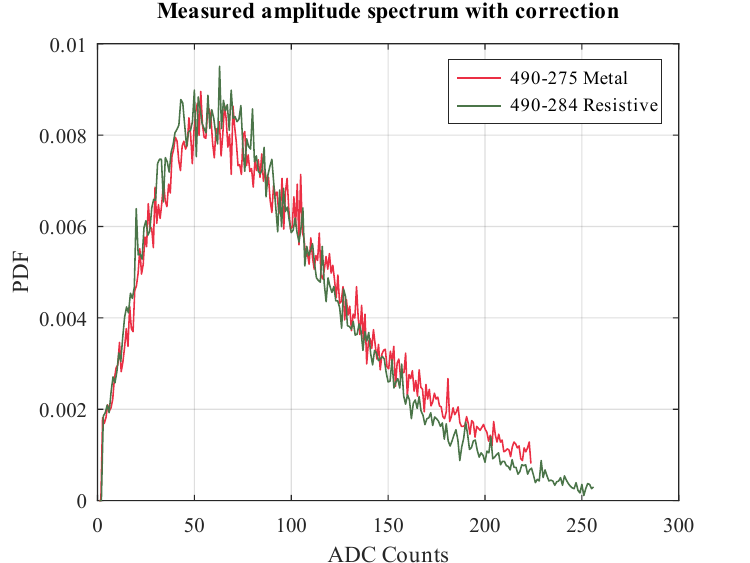}
\caption{Top: Amplitude spectrum of metal and resistive anode detectors measured at the same gain. Bottom: Comparison of the resistive anode detector amplitude spectrum with the metal anode detector spectrum scaled by a signal reduction factor of 0.89.}
\label{fig_amplitude_spectrum}
\end{center}
\end{figure}

Fig. \ref{fig_amplitude_spectrum} (top) shows the recorded amplitude spectrum of the metal and resistive-anode detectors operated at the same effective gain. It can be observed that the metallic detector has a higher mean amplitude value than the resistive-anode detector. This behavior is expected due to the fundamental signal reduction factor, Eq. (\ref{eq_signal_reduction_factor}), that influences the induction of the signal over the dielectric. The spectrum for both the metal and resistive detectors becomes comparable when the spectrum of the former is scaled by the theoretical attenuation factor of 0.89, as shown in Fig. \ref{fig_amplitude_spectrum} (bottom).\\

\begin{figure}[!t]
\begin{center}
\includegraphics[width=.42\textwidth]{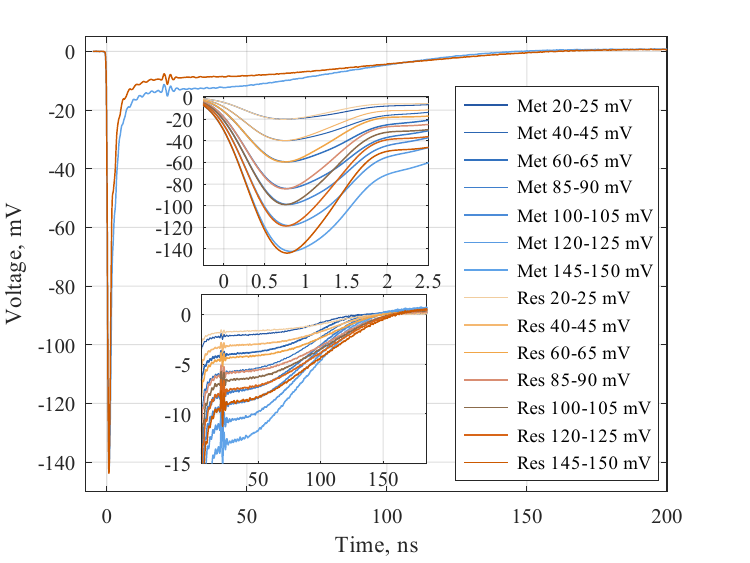}

\caption{Averaged waveforms of metal (blue color) and resistive (orange color) anode detectors with the same electron peak amplitude value between 140 mV and 145 mV. Top sub-figure: the electron peak regions of averaged waveforms with maximum electron peak amplitude value between 20-25, 40-45, 60-65, 85-90, 100-105, 120-125, and 140-145 mV. Bottom sub-figure: Zoomed in ion tail part of the signal corresponding to the e-peak part of the signals shown on top sub-figure.}
\label{fig_wf_comparison_same_gain}
\end{center}
\end{figure}

Signal characteristics of both detectors measured with the same custom made preamplifier were compared at identical detector amplification obtained by the previously described procedure. Events with the same maximum electron-peak amplitude values within the ranges 20 — 25~mV, 40 — 45~mV, 60 — 65~mV, 85 — 90~mV, 100 —105~mV, 120 —125~mV, and 140 — 145~mV were selected for analysis. Fig. \ref{fig_wf_comparison_same_gain} shows averaged waveforms for the metal and resistive-anode detectors for events with amplitude values between 140 mV and 145 mV. The top and bottom subfigures display zoomed-in views of the electron peak and ion tail regions, respectively, for averaged waveforms corresponding to all maximum electron-peak amplitude ranges from 20 to 140 mV.\\

\begin{figure}[!t]
\begin{center}
\includegraphics[width=.42\textwidth]{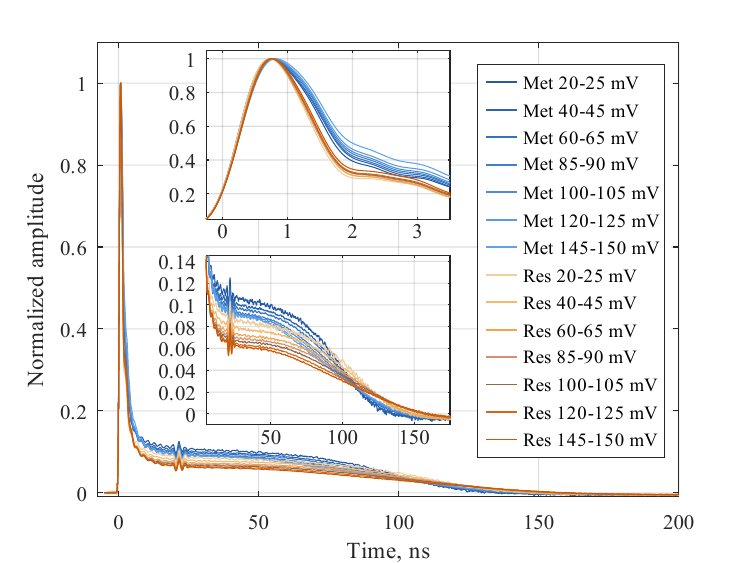}

\caption{Normalized averaged waveforms with the respect to the maximum e-peak amplitude of metal (blue color) and resistive (orange color) anode detectors with the same electron peak amplitude value between 20 mV and 25 mV, 40 mV and 45 mV, 60 mV and 65 mV, 85 mV and 90 mV, 100 mV and 105 mV, 120 mV and 125 mV, and 140 mV and 145 mV. Top sub-figure: Zoomed in electron peak part of the signals. Bottom sub-figure: Zoomed in ion tail part of the signals.}
\label{fig_wf_comparison_same_gain_NORM}
\end{center}
\end{figure}

The shape of the electron-peak leading edge is almost identical for all amplitudes, although the the averaged waveform with electron-peak amplitude value between 140 mV and 145 mV show slightly steeper rising edge for the resistive-anode detector. In contrast, the falling edge of the electron peak is broader for the metal-anode detector at all measured amplitudes, which may be partially attributed to its larger capacitance. Moreover, the ion tail exhibits a larger amplitude and a shorter duration in the metal-anode detector compared to the resistive-anode detector.\\

These similarities and differences in signal shape are shown more clearly in Fig. \ref{fig_wf_comparison_same_gain_NORM}, where all averaged waveforms are normalized to the electron-peak amplitude. The observed differences in the ion tail shape may between metal and resistive detector, in part, be attributed to the delayed component of the induced signal associated with the resistive layer. Furthermore, Fig \ref{fig_wf_comparison_same_gain_NORM} shows that the normalized ion tail plateau reduces with the signal amplitude.  This behavior is observed for both metal and resistive detector. This observed non-linearity could be attributed to the space-charge effect during the drift of the ions; however, this needs to be further investigated. Nevertheless, it is important to note that the electron-peak leading-edge characteristics remain unchanged with the addition of the resistive layer, suggesting that the timing performance of the resistive-anode detector is expected to be comparable to, or better than, that of its metallic counterpart.

\subsection{Detector time response with muon test beam}

\noindent The time response for both types of the single-channel detectors was measured using 150~GeV muons at CERN at the SPS secondary beamline H4. An identical experimental setup to that described in Ref. \cite{utrobicic2025single} was used, consisting of a tracker telescope composed of three triple-GEM detectors and a timing reference and trigger detector based on an MCP-PMT\footnote{Hamamatsu Microchannel plate photomultiplier tube (MCP-PMT) R3809U-50 \url{https://www.hamamatsu.com/jp/en/product/type/R3809U-50/index.html}}. The timing performance of $\varnothing$10 mm PICOSEC Micromegas detectors with metal and resistive-anodes was evaluated using either CsI or DLC photocathodes. The CsI photocathode, with a thickness of 3~nm, was deposited on a 2.38~nm Ti conductive layer on an MgF$_2$ substrate, while the 1.5~nm thick DLC photocathode \cite{lisowska2025photocathode} was deposited directly on MgF$_2$. It was observed that stable operation required a larger drift-gap thickness for the CsI photocathode than for the DLC configuration, namely 120~$\mu$m and 94~$\mu$m, respectively.\\

The PICOSEC Micromegas readout chain consisted of the custom-made preamplifiers and an oscilloscope for signal amplification and digitization. The same oscilloscope was also used to record the serial bitstream containing the event ID information from the tracker, as well as the signal from the timing reference detector (MCP-PMT), which was split twice prior to being connected to the oscilloscope. The oscilloscope was operated with an analog bandwidth of 1 GHz, a sampling rate of 10~GS/s, and a vertical scale of 50~mV/div or 100~mV/div for measurements performed with DLC and CsI photocathodes, respectively.\\

\begin{figure}[!t]
\begin{center}
\includegraphics[width=0.7\columnwidth]{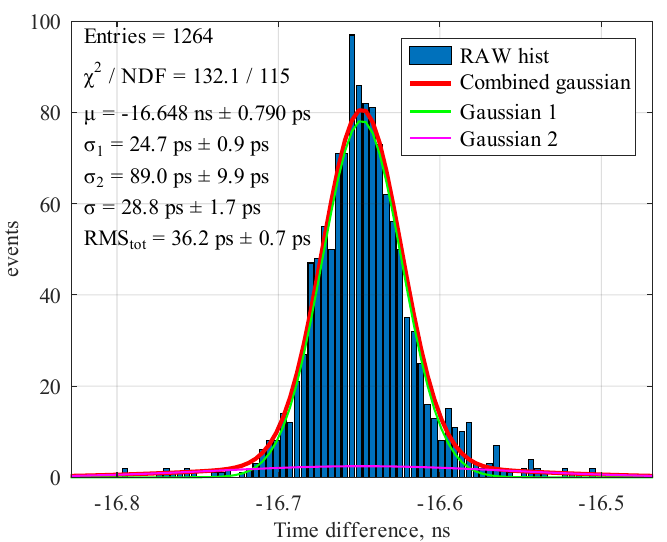}

\vspace{1mm}

\includegraphics[width=0.7\columnwidth]{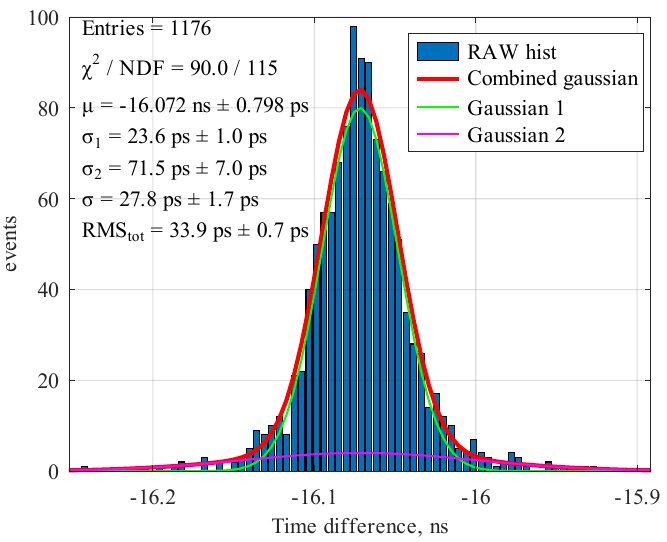}

\caption{Time difference distribution for 150~Gev muons passing within $\varnothing$4~mm pad center. Top: Metal anode detector ($\varnothing$10~mm) with 1.5 nm thick DLC photocathode at cathode and anode voltages of -465~V and 275~V, respectively. Bottom: Resistive anode detector ($\varnothing$10~mm) with 1.5 nm thick DLC photocathode at cathode and anode voltages of -465~V and 285~V, respectively. }
\label{fig_DLC}
\end{center}
\end{figure}

The detector timing performance was analyzed using the same methods as described in Refs. \cite{bortfeldt2018picosec,sohl2020development}.The timing marks for the PICOSEC Micromegas detector and MCP-PMT were obtained at the 20\% constant fraction (CF) from a generalized logistic function fit to the leading edge of the signals. The time resolution was then determined from the distribution of the time differences between the PICOSEC MM and MCP-PMT time marks. Prior to the calculation of the time resolution, several selection criteria were applied to the triggered events in order to define the time-difference distribution: 
\begin{itemize}
    \item A time window cut of $\pm$300 ps around the median time difference was applied to reject outliers.
    \item Additional cuts were imposed on the minimum and maximum signal amplitudes.
    \item Geometrical cuts were applied to the reconstructed event position, retaining only events within a specified diameter around the pad center.
    \item Since the projection of the Cherenkov light cone onto the photocathode surface has a diameter of approximately 6 mm, a tight cut of 4 mm around the pad center was applied to ensure that all contributing photoelectrons were included.
    \item To evaluate the timing performance over nearly the full active area, a wider geometrical cut of 9~mm around the pad center was also applied.
\end{itemize}
The detector time resolution was extracted as the standard deviation of a fit to the time-difference distribution using the sum of two Gaussian functions with a common mean value, following the procedure described in Ref.~\cite{bortfeldt2018picosec}.\\

\begin{figure}[!t]
\begin{center}
\includegraphics[width=0.7\columnwidth]{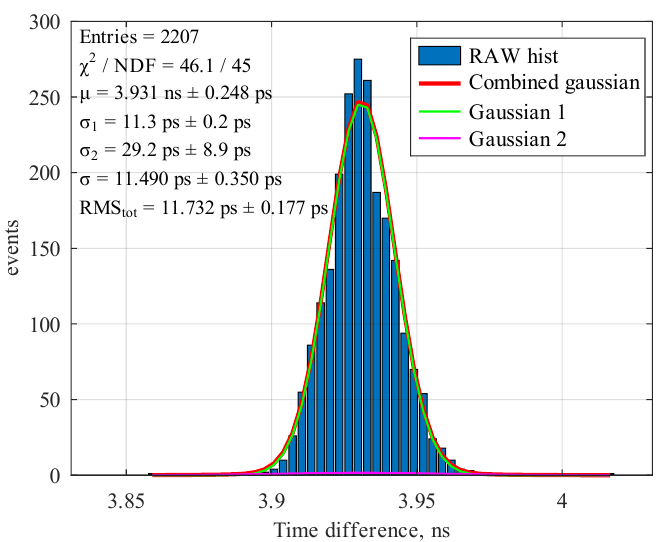}

\vspace{1mm}

\includegraphics[width=0.7\columnwidth]{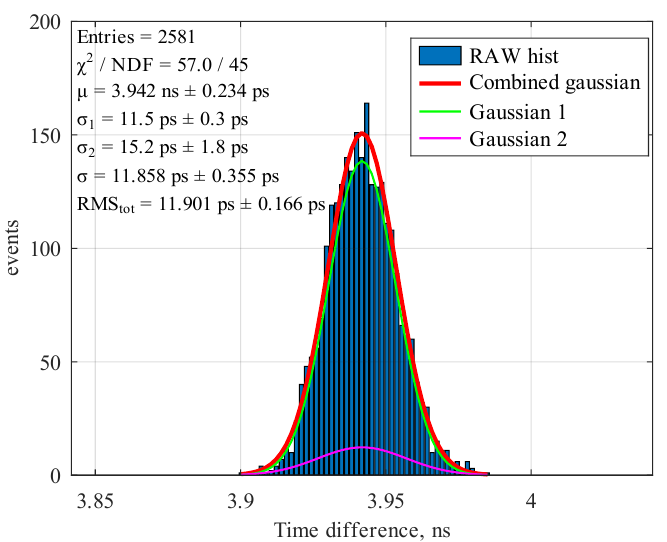}

\caption{Time difference distribution for 150~Gev muons passing within $\varnothing$4~mm pad center for detectors with CsI photocathode. Top: Metal anode detector ($\varnothing$10~mm) operated at cathode and anode voltages of -435~V and 275~V, respectively. Bottom: Resistive anode detector ($\varnothing$10~mm) operated at cathode and anode voltages of -435~V and 284~V, respectively.}
\label{fig_CsI}
\end{center}
\end{figure}
 
Fig. \ref{fig_DLC} (top and bottom) shows the time-difference distributions for 150~GeV muons traversing the central 4~mm-diameter region of the metal and resistive-anode detectors, respectively. An amplitude cut was applied to the PICOSEC MM signals, selecting events with amplitudes between 4~mV and 396~mV. Both detectors were equipped with a 1.5~nm DLC photocathode and a 94~$\mu$m drift gap. The cathode voltage was set to -465 V for both detectors, while the anode voltages were adjusted to 275~V for the metal-anode detector and 285~V for the resistive-anode detector in order to operate at identical effective gain. As a result, the metal and resistive-anode detectors achieved time resolutions of $28.8 \pm 1.7$~ps and $27.8 \pm 0.7$~ps, respectively. When the geometrical cut was expanded to include events over nearly the entire active area ($\varnothing$~9~mm), the time resolution degraded slightly to $32.7 \pm 0.8$~ps and $31.0 \pm 0.9$~ps for the metal and resistive-anode detectors, respectively. The modest degradation observed for the wider geometrical cut is expected, as a fraction of the photoelectrons is produced outside the effective active area of the detector. Nevertheless, in both cases, the two detectors exhibit very similar timing performance. \\

\begin{figure}[!t]
\begin{center}
\includegraphics[width=0.445\columnwidth]{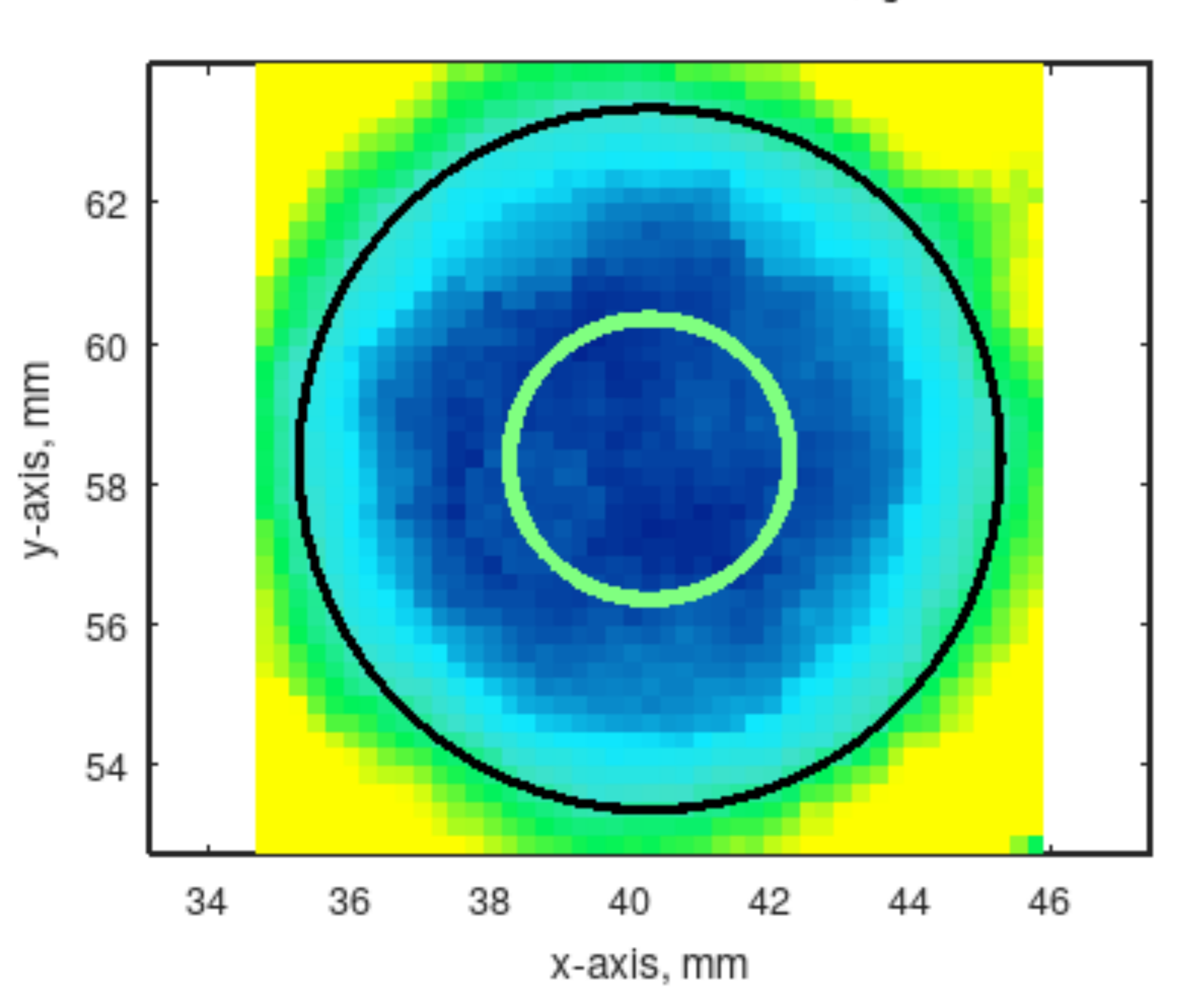}
\includegraphics[width=0.545\columnwidth]{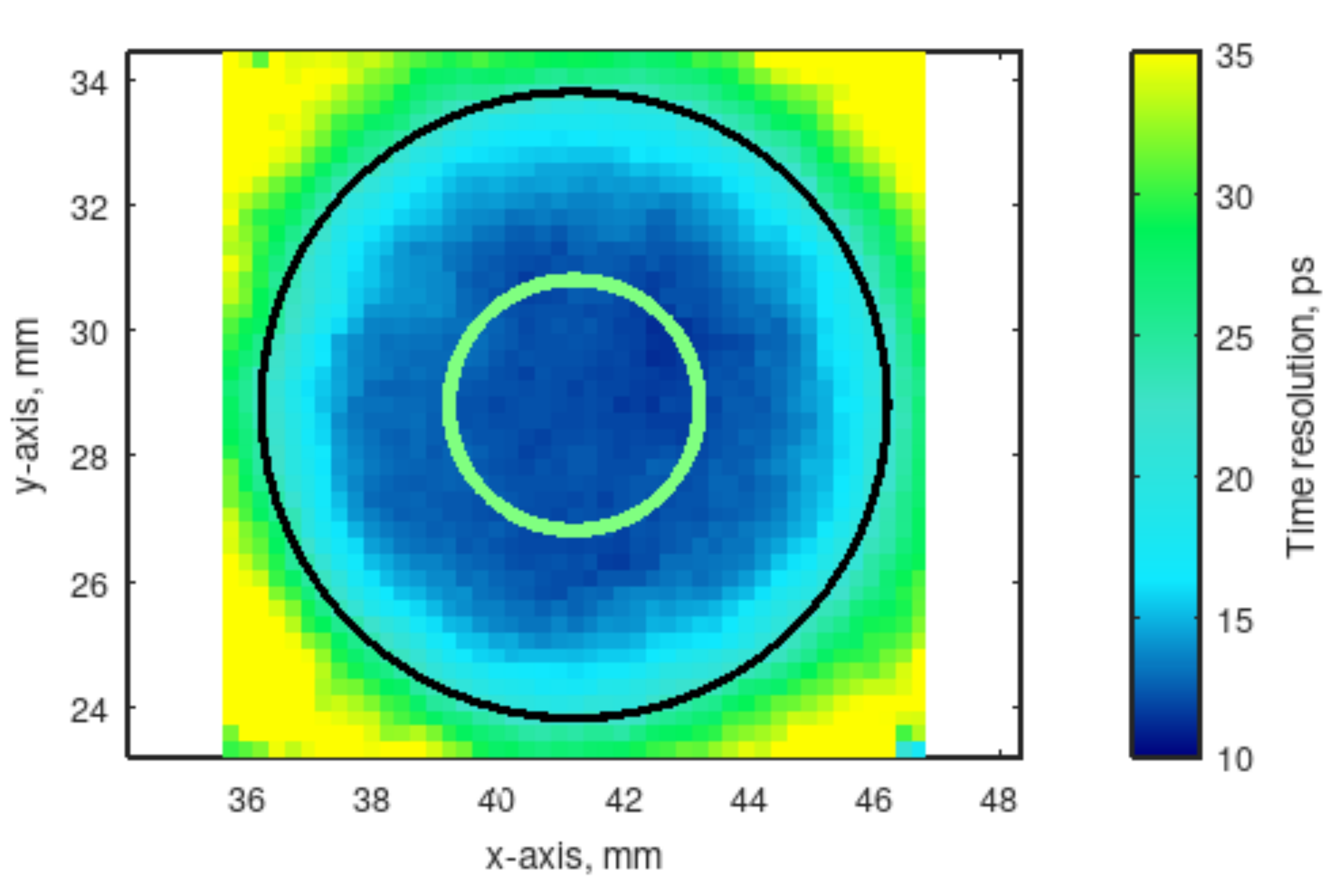}

\caption{Time resolution over the pad area for metal (left) and resistive (right) anode detectors. The black circle indicates the pad’s active area ($\varnothing10~mm$), while the green circle at the centre marks $\varnothing4~mm$ region where the Cherenkov cone from particles passing through this area is fully contained within the pad.}
\label{fig_2D}
\end{center}
\end{figure}

In addition to the measurements using a DLC photocathode, the performance of both detectors was also evaluated using a 3 nm thick CsI photocathode with higher quantum efficiency. For these measurements, both detectors were operated with identical drift-gap thicknesses of 120 µm and at the same cathode voltage of $-435$~V. As before, to achieve identical effective gain, the anode voltages were set to 275~V for the metal-anode detector and 284~V for the resistive-anode detector. Fig. \ref{fig_CsI} shows the time-difference distributions for 150~GeV muons traversing the central 4~mm-diameter region of the metal (top) and resistive-anode (bottom) detectors. The metal-anode detector achieves a time resolution of $11.5 \pm 0.4$~ps, while the resistive-anode detector exhibits a resolution of $11.9 \pm 0.4$~ps, with no statistically significant difference observed between the two. This close agreement indicates that, although the addition of a resistive coating on the insulating layer above the readout pad introduces an inherent reduction in signal induction, the detector maintains excellent timing performance. When the geometrical cut was expanded to include events within a $\varnothing$~9~mm region around the pad center, time resolutions of $13.1 \pm 0.2$ ps and $13.4 \pm 0.5$ ps were obtained for the metal and resistive-anode detectors, respectively. This suggests that, despite the partial loss of photoelectrons outside the active area, the number of detected photoelectrons remains sufficient to sustain high signal amplitudes and good timing resolution, close to that measured at the pad center. This behavior is also illustrated by the two-dimensional maps of the time resolution across the full detector area shown in Fig. \ref{fig_2D} (left and right), where both detector types exhibit uniform and mutually consistent performance over the entire active surface.

\section{Conclusion}
\noindent A resistive-anode implementation of the PICOSEC Micromegas detector was developed and characterized with the aim of improving robustness against discharges and enabling future multi-channel operation while preserving excellent timing performance. A comprehensive study combining analytical modeling, finite-element simulations, and experimental measurements was performed to assess the impact of the resistive layer on rate capability, signal formation, and time resolution.\\

The rate capability of the resistive readout structure was investigated through analytical calculations and FEM simulations of the voltage drop induced by intense irradiation on a finite-size resistive layer. For the single-channel geometry and expected beam conditions at the CERN SPS H4 beam line, surface resistivities around 20~M$\Omega/\Box$ were found to provide an optimal compromise between discharge protection and gain stability. In addition, the influence of the resistive layer on signal integrity was studied using an extended Ramo-Shockley formalism with time-dependent weighting fields and Garfield++ simulations. The contribution of delayed signal components due to the finite conductivity of the DLC layer was quantified, and it was shown that for surface resistivities exceeding approximately 100 k$\Omega/\Box$, the leading edge of the electron peak remains unaffected. While the presence of the insulating layer introduces a fundamental reduction in signal amplitude of about 11\%, laboratory measurements with single photoelectrons confirmed that the electron-peak dynamics and leading-edge characteristics are preserved for the chosen resistivity range.\\

Two resistive-anode prototype detectors with active areas of $\varnothing$10 mm and $\varnothing$15 mm were designed, constructed, and experimentally characterized. Measurements of parasitic capacitances and inductances confirmed the expected changes induced by the resistive layout, which were found to be compatible with fast signal readout. Direct comparisons between metallic and resistive-anode detectors demonstrated consistent signal shapes of the electron-peak under comparable operating conditions, while a difference in both the width of the electron-peak and ion tails were observed. In addition, using a charge sensitive, the 11\% signal attenuation predicted from signal modeling was observed. Finally, beam tests with 150~GeV muons at CERN using both DLC and CsI photocathodes demonstrated that the resistive-anode detector achieves a time resolution fully comparable to that of the metallic-anode configuration. With a CsI photocathode, a resolution of $11.9 \pm 0.4$ ps was measured for the resistive detector, in agreement within uncertainties with the $11.5 \pm 0.4$ ps obtained for the metallic reference. The timing performance was found to be uniform across the active area, confirming that the resistive layer does not degrade the intrinsic time resolution. These results demonstrate that a resistive-anode PICOSEC Micromegas detector can maintain excellent timing performance while providing enhanced operational robustness.
\appendix
\section{Signal amplitude and detector capacitance}

\noindent The signal attenuation caused by the presence of the insulating laminate in the resistive layout can be estimated from the ratio of the detector capacitances of the metal-anode and resistive-anode detectors. For a metallic anode, the capacitance is given by
\begin{equation}
C_{\mathrm{met}} = \varepsilon_0 \frac{A}{g} \, ,
\end{equation}
while for the resistive configuration it becomes
\begin{equation}
C_{\mathrm{res}} = \varepsilon_0 \varepsilon_r \frac{A}{\varepsilon_r g + d} \, .
\end{equation}
Here $A$ denotes the readout-pad area, $g$ the amplification gap size, $\varepsilon_0$ the permittivity of free space, and $\varepsilon_r$ and $d$ the relative permittivity and thickness of the insulating layer, respectively. Taking the ratio yields
\begin{equation}
\frac{C_{\mathrm{res}}}{C_{\mathrm{met}}} = 1 - \frac{d}{d + \varepsilon_r g} \, ,
\label{eq_capacitance}
\end{equation}
which corresponds to the signal reduction factor defined in Eq. \ref{eq_signal_reduction_factor}. As a result, this factor may be determined empirically by measuring the capacitances, provided that parasitic contributions are negligible.
\section*{Acknowledgements}
We thank the CERN-EP-DT-MPT Workshop, particularly Antonio Teixeira, Olivier Pizzirusso, Bertrand Mehl and Rui de Oliveira for the useful discussion and production of MM prototypes with resistive anode.  We acknowledge the financial support of the EP R$\&$D, CERN Strategic Programme on Technologies for Future Experiments; the RD51 Collaboration, in the framework of RD51 common projects; the Cross-Disciplinary Program on Instrumentation and Detection of CEA, the French Alternative Energies and Atomic Energy Commission; the PHENIICS Doctoral School Program of Université Paris-Saclay, France; the Program of National Natural Science Foundation of China (grant number 11935014, 12125505); the COFUND-FP-CERN-2014 program (grant number 665779); the Fundação para a Ciência e a Tecnologia (FCT), Portugal; the Enhanced Eurotalents program (PCOFUND-GA-2013-600382); the US CMS program under DOE contract No. DE-AC02-07CH11359. This material is based upon work supported by the U.S. Department of Energy, Office of Science, Office of Nuclear Physics under contracts DE-AC05-06OR23177.The project is funded by the European Union under the NextGenerationEU Programme.

\vspace{0.5 cm}
\bibliographystyle{elsarticle-num} 
\bibliography{literatura}

\end{document}